\documentstyle[psfig]{mn}

\newif\ifAMStwofonts

\newcommand{\beq}{\begin{equation}}
\newcommand{\eeq}{\end{equation}}
\def\erre{\hbox{\rm\rlap{I}\kern.1em R}}

\newcommand{\be}{\begin{equation}}

\ifoldfss
  \ifCUPmtlplainloaded \else
    \NewTextAlphabet{textbfit} {cmbxti10} {}
    \NewTextAlphabet{textbfss} {cmssbx10} {}
    \NewMathAlphabet{mathbfit} {cmbxti10} {} 
    \NewMathAlphabet{mathbfss} {cmssbx10} {} 
  \fi
  \ifAMStwofonts
    \ifCUPmtlplainloaded \else
      \NewSymbolFont{upmath} {eurm10}
      \NewSymbolFont{AMSa} {msam10}
      \NewMathSymbol{\upi}     {0}{upmath}{19}
      \NewMathSymbol{\umu}     {0}{upmath}{16}
      \NewMathSymbol{\upartial}{0}{upmath}{40}
      \NewMathSymbol{\leqslant}{3}{AMSa}{36}
      \NewMathSymbol{\geqslant}{3}{AMSa}{3E}

      \let\geq=\geqslant 
    \fi
  \fi
\fi 

\ifnfssone
  \newmathalphabet{\mathit}
  \addtoversion{normal}{\mathit}{cmr}{m}{it}
  \addtoversion{bold}{\mathit}{cmr}{bx}{it}
  \newmathalphabet{\mathbfit} 
  \addtoversion{normal}{\mathbfit}{cmr}{bx}{it}
  \addtoversion{bold}{\mathbfit}{cmr}{bx}{it}
  \newmathalphabet{\mathbfss} 
  \addtoversion{normal}{\mathbfss}{cmss}{bx}{n}
  \addtoversion{bold}{\mathbfss}{cmss}{bx}{n}
  \ifAMStwofonts
    \ifCUPmtlplainloaded \else
      \UseAMStwoboldmath
      \makeatletter
      \new@mathgroup\upmath@group
      \define@mathgroup\mv@normal\upmath@group{eur}{m}{n}
      \define@mathgroup\mv@bold\upmath@group{eur}{b}{n}
      \edef\UPM{\hexnumber\upmath@group}
      \new@mathgroup\amsa@group
      \define@mathgroup\mv@normal\amsa@group{msa}{m}{n}
      \define@mathgroup\mv@bold\amsa@group{msa}{m}{n}
      \edef\AMSa{\hexnumber\amsa@group}
      \makeatother
      \mathchardef\upi="0\UPM19
      \mathchardef\umu="0\UPM16
      \mathchardef\upartial="0\UPM40
      \mathchardef\leqslant="3\AMSa36
      \mathchardef\geqslant="3\AMSa3E

      \let\geq=\geqslant 
    \fi
  \fi
\fi 

\ifnfsstwo
  \DeclareMathAlphabet{\mathbfit}{OT1}{cmr}{bx}{it}
  \SetMathAlphabet\mathbfit{bold}{OT1}{cmr}{bx}{it}
  \DeclareMathAlphabet{\mathbfss}{OT1}{cmss}{bx}{n}
  \SetMathAlphabet\mathbfss{bold}{OT1}{cmss}{bx}{n}
  \ifAMStwofonts
    \ifCUPmtlplainloaded \else
      \DeclareSymbolFont{UPM}{U}{eur}{m}{n}
      \SetSymbolFont{UPM}{bold}{U}{eur}{b}{n}
      \DeclareSymbolFont{AMSa}{U}{msa}{m}{n}
      \DeclareMathSymbol{\upi}{0}{UPM}{"19}
      \DeclareMathSymbol{\umu}{0}{UPM}{"16}
      \DeclareMathSymbol{\upartial}{0}{UPM}{"40}
      \DeclareMathSymbol{\leqslant}{3}{AMSa}{"36}
      \DeclareMathSymbol{\geqslant}{3}{AMSa}{"3E}

      \let\geq=\geqslant 
    \fi
  \fi
\fi 

\ifCUPmtlplainloaded \else
  \ifAMStwofonts \else 
    \def\upi{\pi}
    \def\umu{\mu}
    \def\upartial{\partial}
  \fi
\fi

\title{On the clustering phase transition in self-gravitating N-body systems}
\author[M.Cerruti-Sola, P. Cipriani, M.Pettini]
       {Monica Cerruti-Sola$^{1}$,
Piero Cipriani$^{2}$,
and Marco Pettini$^{1}$
\\
$^1$Osservatorio Astrofisico di Arcetri, Largo E. Fermi 5, I-50125 Firenze,
 Italy \\
$^2$Dipartimento di Fisica, Universit\`a ``La Sapienza'', P.le A. Moro, Roma
Italy \\ }

\date{}

\pagerange{\pageref{firstpage}--\pageref{lastpage}}
\pubyear{}

\begin{document}

\maketitle

\label{firstpage}

\begin{abstract}

The thermodynamic behaviour of self-gravitating $N$-body systems has been
worked out by borrowing a standard method from Molecular Dynamics: the
time averages of suitable quantities are numerically computed along the 
dynamical trajectories to yield thermodynamic observables. The link between
dynamics and thermodynamics is made in the microcanonical ensemble of 
statistical mechanics.
The dynamics of self-gravitating $N$-body systems has been computed using 
two different kinds of regularization of the newtonian interaction: 
the usual softening and a truncation of the Fourier expansion series 
of the two-body potential. $N$ particles of
equal masses are constrained in a finite three dimensional volume.
Through the computation of basic thermodynamic observables and of the
equation of state in the $P - V$ plane, new evidence is given of the 
existence of a second order phase transition from a homogeneous
phase to a clustered phase. This corresponds to a crossover from a polytrope 
of index $n=3$, i.e. $p=K V^{-4/3}$, to a perfect gas law $p=K V^{-1}$, as is
shown by the isoenergetic curves on the $P - V$ plane.
The dynamical-microcanonical averages are compared to their corresponding
canonical ensemble averages, obtained through standard Monte Carlo
computations.
A major disagreement is found, because the canonical ensemble seems 
to have completely lost any information about the phase transition. 
The microcanonical ensemble appears as the only reliable statistical framework
to tackle self-gravitating systems.
Finally, our results -- obtained in a ``microscopic'' framework -- are compared
with some existing theoretical predictions -- obtained in a ``macroscopic''
(thermodynamic) framework: qualitative and quantitative agreement is found, 
with an interesting exception.

\end{abstract}


\section{Introduction}
Many astrophysical problems, ranging from cosmology down to the formation 
of planets, demand a knowledge of the physical conditions for the existence 
of stable equilibria as well as for the appearance of instabilities in 
self-gravitating systems, either in the form of large gas clouds or of 
collections of concentrated objects ($N$-body systems). As a consequence, 
on this subject there is a long standing and important tradition of both 
theoretical and numerical investigations.
In view of the aims of the present paper, there are three fundamental 
reference works.
The first one is an old and interesting paper (Bonnor, 1956) where the 
equation of state ($P-V$) was derived for a spherical mass of gas at uniform 
temperature in equilibrium under its own gravitation, and where the modified
Boyle's perfect gas law was related with the onset of a gravitationally
driven instability. The other two papers consider 
$N$ gravitationally interacting particles in a spherical box 
which, under suitable physical conditions, can undergo a catastrophic state
(the Gravothermal Catastrophe) 
characterized by the absence of a maximum of the entropy (Antonov 1962) or,
in a suitable range of parameters, can exist in a clustered phase 
characterized by a negative heat capacity  (Lynden-Bell $\&$ Wood 1968).
Already in this latter paper, then confirmed by subsequent work 
(Aaronson $\&$ Hansen 1972),  it was surmised that -- in analogy with ordinary
thermodynamics -- self-gravitating systems  undergo a phase transition.
This hypothesis was also supported by the theoretical evidence of a 
condensation phenomenon in a system of self-gravitating hard spheres
(Van Kampen 1964).
Also numerical experiments on the $N$-body dynamics supported this analogy
through the evidence of the core-halo formation (Aarseth 1963) or, in
simulations of galaxy clustering (Aarseth 1979), through 
a phenomenology which is reminiscent of a phase transition occurring in 
atomic or molecular systems. More recent evidence of the existence of a phase
transition in gravitational systems has been given for plane and spherical
sheets (Reidl $\&$ Miller 1987; Miller $\&$ Youngkins 1998; Youngkins $\&$
Miller 2000) and for particles constrained on a ring (Sota, Iguchi $\&$ 
Morikawa 2000).

Though the ``negative specific heat paradox'' -- arisen by 
Lynden-Bell $\&$ Wood in their explanation of Antonov's gravothermal 
catastrophe -- seems to faint the analogy with laboratory systems, 
Hertel and Thirring (Hertel $\&$ Thirring 1971) showed that a negative 
specific heat -- which is strictly forbidden in the canonical ensemble of
statistical mechanics --  
can be legally found in the microcanonical ensemble as a consequence of 
the breakdown of ensemble equivalence; this ensemble
inequivalence has been later
confirmed for a simplified model of gravitationally interacting objects 
(Lynden-Bell $\&$ Lynden-Bell 1977)\footnote{The inequivalence of statistical
ensembles is not limited to gravitational systems, but it occurs whenever the
range of the potential is comparable with the size of a system (Cipriani $\&$
Pettini 2000) and has recently attracted a lot of interest (Gross 1997).}.

All the results of the three above quoted papers were worked out in a 
{\it macroscopic} thermodynamic framework. Thus we can wonder if and how such
a macroscopic phenomenology can be retrieved in the framework of a
{\it microscopic} dynamical description. In other words, in the spirit  of
statistical mechanics, we can try to link the dynamics of the elementary
constituents of a self-gravitating system with its large scale thermodynamics.
(Throughout this paper we use ``macroscopic'' and ``microscopic'' in a 
relative sense, thus 
the microscopic level can be that of individual atoms or molecules of
a gas cloud, as well as that of stars in a galaxy). This is actually one of 
the aims of the present paper, where we address the problem of the statistical
mechanical treatment of self-gravitating systems in order to understand
whether the known thermodynamics can be derived from purely gravitational
dynamics. The other aim of the present paper, closely related to the previous 
one, is to tackle the clustering instability, i.e. the breakdown of the
homogeneous phase into a core-halo phase, in the attempt to describe it in
analogy with the other -- better understood -- laboratory phase transitions.

The paper is organized as follows. In Section \ref{N-body} we briefly discuss 
why a-priori a statistical mechanical treatment of gravitational $N$-body
systems runs into difficulties and, on the other hand, why these systems can 
be  considered {\it bona fide} ergodic after suitable regularization; 
moreover,
we present an unconventional regularization strategy of the exact Hamiltonian
(transformed into a ``Fourier-truncated'' Hamiltonian),  
leading to models that do not afford any practical advantage in the numerical 
treatment of the $N$-body dynamics but that are of great prospective interest 
for its analytic treatment in a statistical mechanical context and thus to 
apply
concepts and methods that already proved powerful in the study of Hamiltonian 
dynamics. In Section \ref{numerics}, after a quick glance at the integration
method and parameters of the equations of motion and at the MonteCarlo
simulations of the canonical ensemble averages, we show how thermodynamics
can be worked out through dynamics with the aid of microcanonical statistical
ensemble and by borrowing from Molecular Dynamics useful formulae already
in the literature. We report the caloric curves (temperature
{\it vs} energy), the specific heat and an order parameter {\it vs} energy 
(suitably scaled with $N$), and a comparison among the outcomes of the
simulations obtained with the Fourier-truncated Hamiltonian and those
obtained with the 
standardly softened Hamiltonian. A somewhat familiar phenomenology in the
numerical study of phase transitions is found mainly through the order 
parameter. A strikingly strong ensemble inequivalence is found: the
canonical MonteCarlo averages do not keep even the slightest trace of the
change of the specific heat from positive to negative and of the phase 
transition.
Having confirmed the a-priori expected ensemble inequivalence, we discuss
the reasons to consider the microcanonical ensemble as the good representative
statistical ensemble for self-gravitating $N$-body systems and we present the 
dynamically worked out $P-V$ equation of state. 
In Section \ref{theory} we compare our results with their theoretical
counterparts obtained within a thermodynamic macroscopic approach. Qualitative
and quantitative agreement is found and everything fits into a coherent
scenario, though with a remarkable difference between the clustering 
transition
and the gravothermal catastrophe. Finally, in Section \ref{concl} some
conclusions are drawn. In Appendix A some basic definitions and concepts 
of the
Gibbsian ensemble formulation of statistical mechanics are briefly recalled.
In Appendix B a technical detail about the mean-field approximation is 
sketched.
\section{Self-gravitating N-body systems}
\label{N-body}
\medskip
In principle star clusters, galaxies, clusters of galaxies seem to naturally 
call for a statistical description of their dynamical behaviour. 
However, as the existence of negative specific heats reveals, there are
some difficulties due to  
the very special nature of gravitational interaction.
The  gravitational interaction is always attractive, unscreened and of infinite
range, therefore it is not stable, i.e. the 
potential energy $U({\bf r}_1,\dots,{\bf r}_N)$ of $N$ 
gravitationally interacting 
masses does not fulfil the condition $U({\bf r}_1,\dots,{\bf r}_N)\geq -N B$,
with $B$ a positive constant, whence the so-called lack of saturation:  
in the $N\rightarrow\infty$ limit the binding energy per particle and
the free energy per particle diverge. This is
also referred to as a breakdown of extensivity of these fundamental physical
quantities, at variance with what is familiar for ordinary laboratory systems.
Moreover, from a rigorous mathematical viewpoint, there is no equilibrium
ground state: as $N$ increases, at $N/V=const$, the energy per particle
increases with $N$ and cannot be defined in the thermodynamic limit
\footnote{Though, from a physical point of view, the thermodynamic limit 
does not exist and has no meaning for self-gravitating systems.}.
The absence of an equilibrium state  means that a gravitational system does 
not behave thermodynamically because standard thermodynamics does not apply to
evolving systems.
However, a thermodynamic description is still possible for self-gravitating 
$N$-body systems provided that they are not in a strongly unstable phase 
(Saslaw 1985), and in the case of {\it slowly} evolving system
we can resort to statistical mechanics.

Moreover, statistical mechanics gives a {\it correct} thermodynamics if  
the thermodynamic potentials are extensive quantities.
In a self-gravitating system only entropy in the
microcanonical ensemble -- at least on finite time scales and if the system 
is slowly evolving -- is an extensive thermodynamic potential\footnote{From
the fundamental relation $1/T=\partial S/\partial E$, being $T$ proportional
to the kinetic energy per particle and making use of the virial theorem,
$NT$ and $E$ must have the same $N$ dependence hence $S$ is extensive.};
therefore, the results obtained in other ensembles
(grancanonical and canonical) are a-priori expected to be in disagreement 
with the results obtained in the microcanonical ensemble.  

\smallskip
In what follows, systems of $N$ gravitationally 
interacting point masses will be considered that are described by the 
Hamiltonian function 
\begin{equation}
H = \sum_{i=1}^N\frac{1}{2m_i}\left( p_{xi}^2 + p_{yi}^2 + p_{zi}^2\right) -
\frac{G}{2}\sum_{i,j=1}^N(1-\delta_{ij})\frac{m_i m_j}
{\vert{\bf r}_i-{\bf r}_j\vert}
\label{Hvera}
\end{equation}
where ${\bf r}_i\equiv (x_i, y_i, z_i)$.
For the sake of simplicity we shall put $G=1$ and $m_i=1,~i=1,\dots , N$.
We remark that this choice is convenient to keep the statistical and 
thermodynamic properties unaltered (e.g. the non-extensivity), at variance 
with what is implied by other possible choices (Heggie $\&$ Mathieu 1986).
\subsection{Dynamics and statistics}
Any physical phenomenon occurring in the system described by Hamiltonian
(\ref{Hvera}) must have its origin in the dynamics. 
With the exception of those systems that can be treated with the methods of
celestial mechanics (mainly perturbation theory), the dynamics of generic 
$N$-body systems can be worked out only through numerical simulations. The 
numeric approach, which is continually improved (Aarseth 1999; Meylan $\&$ 
Heggie 1997, and references quoted in these papers), provides the 
``experimental'' counterpart of the theoretical investigation of these systems.
 
In order to describe the physics of $N$-body systems through 
a few relevant macroscopic observables, dynamics must fulfil the requirements 
of ergodicity and
mixing to justify the application of statistical mechanics. 
It is in general impossible for Hamiltonian systems of physical interest to 
ascertain whether ergodicity and mixing are rigorously verified. However, 
a generic non-integrable and chaotic Hamiltonian system with a large number 
$N$ of degrees of freedom can be considered ergodic and mixing, at least in 
a physical sense. In fact, after 
a famous theorem due to Poincar\'e and Fermi (Poincar\'e 1892; 
Fermi 1923a,b), generic systems with three or more degrees of freedom are
not integrable, i.e. there are no nontrivial invariants of motion besides 
total energy. Only global invariants (like total momentum and angular 
momentum), due to global symmetries of the 
Hamiltonian (\ref{Hvera}) (as space translations and rotations), can exist. 
Thus, once the initial condition of a gravitational $N$-body system is 
assigned, the representative point of the system moves on a $6N-10$ 
dimensional hypersurface of its $6N$-dimensional phase space. 
The lack of nontrivial integrals of motion (i.e. not related to global 
symmetries) entails that all the $6N-10$ dimensional hypersurface of phase 
space is accessible to a trajectory issuing from any initial condition. 
A possible source of non-ergodicity would be the coexistence 
of regular and chaotic motions:  observed for the first time in the 
H\'enon-Heiles model (H\'enon $\&$ Heiles 1964), it is in general implied by 
the Kolmogorov-Arnold-Moser (KAM) theorem (Thirring 1978), which, however, 
has no practical relevance at large $N$ 
\footnote{KAM theorem requires extremely tiny deviations 
from integrability to imply the existence of sizeable regions of regular 
motions in phase space, moreover these deviations from integrability cannot
exceed a threshold value which drops to zero exponentially fast with $N$ 
(Thirring 1978; Casetti et al. 1999) .}. 
Thus, self-gravitating $N$-body systems, after 
some suitable regularization to make finite the phase space volume, can be
considered ergodic, so that time averages of physically relevant observables 
can be replaced with suitable statistical averages computed on a given 
$6N-10$ dimensional hypersurface of phase space.

The dynamical instability of the gravitational $N$-body systems 
(Miller 1964) implied one among the first numerical evidences of the existence
of deterministic chaos in Hamiltonian dynamics. Since then, further numerical 
evidence of the existence of chaos in the self-gravitating systems has been 
provided by several authors (Kandrup 1990; Quinlan $\&$ Tremaine 1992; 
Goodman, Heggie $\&$ Hut 1993; Kandrup, Mahon $\&$ Smith 1994, and references 
therein; Cipriani $\&$ Pucacco 1994; Cerruti-Sola $\&$ Pettini 1995; 
Cipriani $\&$ Di Bari 1998; Miller 1999).
On the other hand, chaotic dynamics in a many dimensional phase space
implies a {\it bona fide} phase mixing (Casetti et al. 1999), which means that
time averages of physical observables converge to their statistical 
counterparts in a {\it finite} time
(whereas ergodicity implies an infinite time convergence). 

Therefore, the use of microcanonical statistical mechanics is naturally
motivated by the reasonably good ergodicity and mixing properties of the 
dynamics of regularized self-gravitating $N$-body systems (Cipriani 
$\&$ Pettini 2000). 
Taking into 
account the conservation of energy, of center of mass position and momentum,
and of angular momentum, the microcanonical volume in phase space reads
\begin{equation}
\omega_N(E) =  
\int\prod_{i=1}^N  d{\bf r}_i d{\bf p}_i 
\delta (H({\bf p},{\bf r}) - E) 
\label{microangmom}
\end{equation}
\[
~~~~\delta^{(3)}(\sum_i {\bf r}_i\wedge{\bf p}_i - {\bf L}) 
\delta^{(3)}(\sum_i {\bf p}_i - {\bf P})
\delta^{(3)}(\sum_i {\bf r}_i - {\bf R})~.
\]
In principle, by means of $\omega_N(E)$,  we can compute statistical averages
of any physical observable defined through a function $A({\bf p},{\bf r})$,
and we can also compute the thermodynamics of a self-gravitating $N$-body 
system, the link being made by the entropy defined as $S=k_B \log \omega_N(E)$;
$k_B$ is the Boltzmann constant (see Appendix A).

\subsection{Regularized N-body Hamiltonians}

Let us now consider a system constrained in a finite volume. At variance with
the customary choice of a spherical box (Bonnor 1956; Antonov 1962; 
Lynden-Bell $\&$ Wood 1968), let us consider a cubic box of 
side length equal to $L$. The reason for such a choice is to explicitly break
the rotational invariance of the Hamiltonian (\ref{Hvera}) and, in so doing, 
the microcanonical volume $\omega_N(E)$ simplifies to\footnote{In fact, 
both periodic conditions and reflecting boundary conditions destroy
the conservation of {\bf P} and {\bf R}.} 
\begin{equation}
\omega_N(E) =  
\int\prod_{i=1}^N  d{\bf r}_i d{\bf p}_i \ 
\delta (H({\bf p},{\bf r}) - E)~,
\label{micro}
\end{equation}
and thus we can borrow from the existing literature the analytic expressions, 
derived using $\omega_N(E)$ of Eq.(\ref{micro}),  
of some basic thermodynamic observables that are then used in our numerical
computations. Working out anew the same analytic expressions using 
$\omega_N(E)$ of
Eq.(\ref{microangmom}) is a non trivial task beyond the aims of our present
investigation. However, this is not a severe restriction if we refer to
almost non-rotating systems whose angular momentum, even if conserved, is 
negligible. The cubic box, allowing fluctuations of the total angular momentum,
is thus equivalent to considering a whole ensemble of almost vanishing angular 
momentum systems.

Even though the assumption of a confining box could seem unphysical, it is a
simple way of idealizing different physical aspects which depend upon the 
chosen boundary conditions. In fact, the assumption of periodic boundary 
conditions is as if we took a fragment out 
of the bulk of a large system where -- in the average -- the number of particles
remains constant and small local density and energy fluctuations of the 
subsystem take place. In this case when a
mass exits the box in a given direction, another mass with the same energy 
enters the volume from the opposite
side to keep constant the energy and the number of particles.
The assumption of reflecting boundary conditions amounts to mimic the 
presence of a halo of diffuse matter whose gravitational
potential field would act to confine the system. Both assumptions about the 
geometric constraints of the system, besides the
explicit breaking of rotational symmetry, also guarantee the finiteness of the
configuration space volume over which the integration in Eq. (\ref{micro}) is
performed. 

In order to guarantee also the boundedness in momentum space, so 
that the whole integral in Eq.(\ref{micro}) extends over a finite region of 
phase space, the two-body interaction potential must be regularized.

We adopted two different kinds of regularization. 
The first one is the standard
softening, adopted in numerical simulations, with the replacement (e.g. see
Binney $\&$ Tremaine 1987)
\begin{equation}
\frac{1}{\vert{\bf r}_i - {\bf r}_j\vert}~\longrightarrow~
\frac{1}{\sqrt{({\bf r}_i - {\bf r}_j)^2 + \eta^2} }~,
\label{soften}
\end{equation}
where $\eta$ is a small softening parameter that bounds below the interaction
potential and in so doing prevents the occurrence of arbitrarily large values 
of the momenta.
This regularization is local in space. 

The second regularization is nonlocal in space. It makes use of the Fourier 
representation of the Green function $G({\bf r}-{\bf r}^\prime)$ for the 
Poisson equation
\begin{equation}
\nabla^2 G({\bf r}-{\bf r}^\prime) = - 4\pi \delta ({\bf r}-{\bf r}^\prime)
\label{poisson}
\end{equation}
in a cubic box of side $L$. In fact, one has 
$G({\bf r}-{\bf r}^\prime)=1/\vert
{\bf r}-{\bf r}^\prime\vert$ with the following Fourier development  
(Jackson 1975) 
\begin{eqnarray}
\hskip -.6truecm
 &&\frac{1}{\vert {\bf r}-{\bf r}^\prime\vert} = \hfill\nonumber\\
\hskip -.6truecm
 & & \frac{32}{\pi L}\sum_{l,m,n=1}
^\infty \frac{\sin ( k_l x) \sin (k_m y) \sin (k_n z)
 \sin ( k_l x^\prime ) \sin (k_m y^\prime ) \sin (k_n z^\prime )}
{ (l^2+m^2+n^2)}~,\hfill\nonumber\\
\hskip -.6truecm
&&
\label{green}
\end{eqnarray}
where $k_l=\pi\ l /L$, $k_m=\pi\ m /L$ and $k_n=\pi\ n /L$. Thus, Hamiltonian 
(\ref{Hvera}) can be exactly rewritten as
\[
H = \sum_{i=1}^N\frac{1}{2m_i}\left( p_{xi}^2 + p_{yi}^2 + p_{zi}^2\right) -
\frac{16\ G}{\pi L}\sum_{i,j=1}^N(1-\delta_{ij})m_i m_j 
\]
\[
\sum_{l,m,n=1}
^\infty \frac{\sin ( k_l x_i)\sin (k_m y_i)\sin (k_n z_i)
\sin ( k_l x_j )\sin (k_m y_j )\sin (k_n z_j )}{ (l^2+m^2+n^2)}~,
\]
\begin{equation}
\label{Hfourier}
\end{equation}
\vskip -.5truecm
which, in an arbitrarily large volume, is completely equivalent to
(\ref{Hvera}).

In order to regularize the two-body potential in Hamiltonian
(\ref{Hvera}), one can truncate the Fourier expansion in Hamiltonian 
(\ref{Hfourier}) by summing $l,m,n$ from $1$ up to a finite number 
${\cal N}_w$. 

The two regularizations are a-priori inequivalent:
the former pertains events (close encounters) localised in real space, 
whereas the latter makes
use of collective coordinates (the Fourier modes) which are not localised in
real space. By truncating the Fourier expansion in Hamiltonian (\ref{Hfourier})
we neglect all the dynamical details occurring at length scales smaller than
the smallest spatial wavelength in the expansion. Loosely speaking, this is
reminiscent of standard methods in statistical mechanics, mainly in the 
context of the renormalisation group theory, where
relevant and irrelevant degrees of freedom at a given length scale are 
Fourier modes with wavelengths above and below some cutoff respectively.
In order to ascertain to what extent a truncated model still retains some 
physically relevant feature of the exact $N$-body system (\ref{Hvera}), it
is necessary to compare the outcomes of different truncations, i.e.
different ${\cal N}_w$, and to make a comparison between the results obtained
by simulating the dynamics associated with Hamiltonian (\ref{Hvera}) and with
a truncated version of the  Hamiltonian (\ref{Hfourier}).

There are different reasons for making numerically heavier the already heavy
$N$-body problem. First of all, the spatial coarse-graining 
 considerably lessens the dramatic effect that close encounters have on a
reliable numerical computation of Lyapunov exponents (though they are not 
reported here), as well as on other
observables of dynamical relevance; the frequency and the effect of close 
encounters in numerical simulations of $N$-body systems are in this respect
too much enhanced in comparison with the physical reality of practically
collisionless systems as galaxies, cluster of galaxies and, to a lesser extent,
star clusters. 
Second, the approach to equilibrium, as well as many other
non-trivial dynamical properties, are by far better revealed by the use of 
collective coordinates, as a long experience in a different context has widely
witnessed (Casetti et al. 1999). 
Third, truncated Hamiltonians out of (\ref{Hfourier}) have the interesting 
property of naturally allowing a mean-field-like decoupling of the degrees of 
freedom which is very interesting in view of analytical computations of both
statistical mechanical properties and of chaotic properties of the dynamics
in the framework of a geometric theory of Hamiltonian chaos (Casetti et al. 
2000); moreover, such a mean-field-like representation of the Fourier
truncated Hamiltonians leads to the definition of an order parameter that is  
useful to detect the occurrence of a phase transition.
The regularization helps to smooth the local fluctuations on small spatial 
scales which do not significantly affect the macroscopic behaviour of the 
system, but cause a noisy variation of the averages on the time scales of 
interest.
Finally, a wealth of models could be generated, even by retaining
very few modes, which could reveal different aspects of the astonishingly
rich dynamics of the exact $N$-body problem.

A generic truncated model Hamiltonian reads
\[
{\overline H} 
= \sum_{i=1}^N\frac{1}{2}\left( p_{xi}^2 + p_{yi}^2 + p_{zi}^2\right) -
\frac{16}{\pi L}\sum_{i,j=1}^N(1-\delta_{ij})
\]
\[
\sum_{l,m,n=1}
^{{\cal N}_w} \frac{\sin ( k_l x_i)\sin (k_m y_i)\sin (k_n z_i)
\sin ( k_l x_j )\sin (k_m y_j )\sin (k_n z_j )}{ (l^2+m^2+n^2)}
\]
\begin{equation}
\label{Hbar}
\end{equation}
where we have chosen $G=1$ and $m_i=1,~i=1,\dots ,N$. From the associated 
Lagrangian function -- being $p_{xi}=\dot x_i$,  $p_{yi}=\dot y_i$,  
$p_{zi}=\dot z_i$ -- 
the following equations of motion are derived
\begin{eqnarray}
{\ddot x}_i & = &\frac{32}{L^2}\sum_{l,m,n=1}^{{\cal N}_w}
\frac{l S_{l,m,n}^{(i)}}
{ (l^2+m^2+n^2)} \cos ( k_l x_i) \sin (k_m y_i) \sin (k_n z_i)\hfill\nonumber\\
{\ddot y}_i & = &\frac{32}{L^2}\sum_{l,m,n=1}^{{\cal N}_w}
\frac{m S_{l,m,n}^{(i)}}
 {(l^2+m^2+n^2)} \sin ( k_l x_i)\cos (k_m y_i) \sin (k_n z_i)\hfill\nonumber\\
{\ddot z}_i & = &\frac{32}{L^2}\sum_{l,m,n=1}^{{\cal N}_w}
\frac{n S_{l,m,n}^{(i)}}
{ (l^2+m^2+n^2)}\sin ( k_l x_i) \sin (k_m y_i)\cos (k_n z_i)\hfill\nonumber\\
&& 
\label{eqmotion}
\end{eqnarray}
$i=1,\dots ,N$, and where we have introduced
$S_{l,m,n}^{(i)}= S_{l,m,n}-\sin ( k_l x_i) \sin (k_m y_i) 
\sin (k_n z_i)$, with $S_{l,m,n} = \sum_{j=1}^N\sin ( k_l x_j) \sin (k_m y_j) 
\sin (k_n z_j)$, to put in evidence that the same quantities $S_{l,m,n}$
enter  all the equations of motion, what obviously simplifies the numerical
computations.

\section{Numerical results}
\label{numerics}
The phase space trajectories of an Hamiltonian system are constrained on a 
constant energy surface in phase space; therefore, time averages computed 
along the numerical
solutions of the equations of motion, of both the Fourier-truncated system and
of the softened version of the Hamiltonian (\ref{Hvera}), are generically
expected to converge to microcanonical ensemble averages\footnote{Both 
regularizations do not qualitatively alter the chaoticity of the dynamics.} 
(see Appendix A).
Thus, in order to work out the thermodynamics of self-gravitating $N$-body 
systems, we borrow from Molecular Dynamics the formulae that link microscopic 
dynamics with macroscopic thermodynamics (Pearson, Halicioglu $\&$ 
Tiller 1985).

At variance, the numerical estimate of canonical ensemble averages requires
to construct suitable random markovian walks in the full phase space. Along
such random trajectories (no longer constrained on any energy surface), the
averages of  physical observables converge to canonical ensemble averages 
provided that the recipe to generate the random walk is that of the so-called
Metropolis {\it importance sampling} of the canonical ensemble weight. 

\subsection{Numerical integration}
The numerical integration of the equations of motion (\ref{eqmotion}) and of
the equations of motion derived from the Hamiltonian (\ref{Hvera}) with the
replacement (\ref{soften}) has been performed by means of a symplectic 
algorithm (McLahlan $\&$ Atela 1992). 
Some runs to check the reliability of the results have been performed
using also a bilateral scheme (Casetti 1995).
Symplectic integrators update the coordinates 
and momenta of an Hamiltonian system through a canonical transformation of
variables;
for this reason, symplectic algorithms ensure a faithful representation of an
Hamiltonian flow because, in addition to the conservation of phase space 
volumes and of the energy, they guarantee the conservation of all the 
Poincar\'e integral invariants of a system. Actually, there are interpolation 
theorems (Moser 1968; Benettin $\&$ Giorgilli 1994) stating that the 
numerical flows obtained
through symplectic algorithms can be made as close as we may wish to the
true flow of a given Hamiltonian.
Though locally other integration schemes  can give more precise results in 
presence of close encounters by using different time steps for 
individual particles (Aarseth 1985), the non-symplectic nature of these 
algorithms might 
a-priori somewhat alter the frequency with which different regions of phase 
space have to be visited by an ergodic dynamics, at least for very long runs. 
On the other hand, we are just interested in long runs, so that time averages 
of the chosen observables display a good stabilization, and, in order to safely
replace microcanonical ensemble averages with time averages, dynamics has to 
properly sample the phase space.

The dynamics of the direct $N$-body system has been numerically computed using 
integration time steps
$\Delta t$ ranging in the interval $(5\cdot 10 ^{-5} - 10 ^{-3})$:
the relative variations $\Delta E/E$ of the total energy were
in the range $(10^{-10}, 10^{-4})$ on integration runs of $(10^6, 10^7)$
time steps. The softening parameter $\eta$ has been set to $0.01$ and scaled as
$\eta = \eta d$, where $d = D/N^{1/3}$, with $D= min(N^2/E,L)$.

For what concerns the Fourier-truncated model, computations have been done
with ${\cal N}_w= 5, 7, 10$, 
i.e. a total number of modes equal to $125, 343$ and $1000$ respectively.
The initial conditions on the particle coordinates $(x_i,y_i,z_i)$ have been 
chosen at random with a uniform distribution in the interval $(0, L)$.
The initial momenta have been chosen at random with a gaussian
distibution of zero mean and a suitable variance to approximately match the
desired initial value of the total energy. An opportune velocity rescaling
allowed to precisely fix the initial value of the total energy.
With integration time steps $\Delta t$ ranging in the interval $(0.01, 0.001)$,
we got relative variations $\Delta E/E$ of the total energy
in the range $(10^{-8}, 10^{-6})$ with zero mean, i.e. without any drift,
and on  long integration runs of $10^6$ time steps.
For $E<0$, the system has been then let evolve for about $3 t_D$, where the 
dynamical time is defined as $t_D\propto (N/\vert E/N\vert^{3/2})$ 
(this makes the system virialize); for $E>0$, we let the system evolve until 
it attains  a stationary state of dynamical equilibrium between kinetic and 
potential energies.

By varying the side $L$ of the box, the volume $V$ has been varied.
With both periodic and reflecting boundary conditions, the initial velocity of 
the center of mass has been set equal to zero. The initial total angular 
momentum can be made very small, but then it fluctuates because of the explicit
breaking of rotational symmetry. The larger $N$, the smaller such fluctuations
can be.

\subsubsection{The N-scaling of the results}
\label{Nscaling}
The number $N$ of interacting bodies has been varied in the range $25 - 500$,
with some checks up to $N=2000$.
In the numerical simulations of extensive Hamiltonian systems, i.e. with
short-range interactions so that energy and  other physical observables are
additive, the comparison among the results obtained by varying $N$ is naturally
made through densities: the values of the observables  divided by $N$.

In what follows,  we shall mainly vary the
energy $E$ by keeping the ratio $\varrho =N/V$ constant.
From a dimensional point of view, the potential energy $U=-\sum_{i,j}^{1,...,N}
1/r_{ij}$ is $[U]=N^2/L$, and as $L=V^{1/3}=(N/\varrho)^{1/3}$  
($\varrho$ is constant), we get $U\sim \varrho^{1/3} N^{5/3}$. 
As we shall see in the following, this actually suggests the correct way of 
scaling the results obtained for different values of $N$.

\subsection{MonteCarlo computations}
For standard Hamiltonians $H=\sum_i \frac{1}{2}p_i^2 + U(q)$, the weight
$e^{-\beta H}$ splits into a multidimensional gaussian, originated by the 
kinetic energy term, and into a configurational part $e^{-\beta U(q)}$; 
$\beta$ is the inverse of the temperature.
Only this latter part is dealt with by the MonteCarlo algorithm.
The system under consideration has to be at equilibrium
so that the transition probabilities $W(1\rightarrow 2)$ and 
$W(2\rightarrow 1)$ between any two configurations ``$1$'' and ``$2$''
satisfy the condition $W(1\rightarrow 2) = W(2\rightarrow 1)$, the so-called
detailed balance. A MonteCarlo move consists of a random update of the
coordinates. If the system lowers its
energy with the proposed configuration update, then the new configuration 
is accepted, otherwise the
transition probability $W(\{q_i\}\rightarrow\{q^\prime_i\})={\cal N} \exp \{
-\beta [U(q^\prime_i)- U(q_i)]\}$, where ${\cal N}$ is the normalization
factor, is compared with a random number $\zeta$ uniformly distributed in the
interval $[0,1]$; if $W >\zeta$ then the new configuration is accepted,
otherwise the old configuration is counted once more. This is the essential
of the Metropolis importance sampling algorithm (Binder 1979). The average 
acceptance rate is usually kept not far from $50$ per cent by adjusting the 
mean
variation of the coordinates at each update proposal. The MonteCarlo estimate
of the canonical average of an observable $A$ is then simply given by
\[
\langle A\rangle =\frac{1}{N_{MC}}\sum_{\{q_i\}} A(\{q_i\})~,
\]
the sum being carried over the $N_{MC}$ accepted configurations $\{q_i\}$.

\subsection{Dynamical analysis of thermodynamical observables}
In deriving the thermodynamics of  self-gravitating $N$-body systems, one
of our aims is to address the
problem of the existence and of the characterization of the clustering phase 
transition. It is not out of place to remind here that
the basic thermodynamical phenomenology of a phase transition is 
characterized by the sudden qualitative change of a macroscopic system
when its temperature varies across a critical value $T_c$. This sudden change
is qualitatively due to collective microscopic behaviours and  
is quantitatively reflected by the singular\footnote{True singularities are
possible only in the $N\rightarrow\infty$ limit. Here ``singular'' means
that the larger $N$ the sharper is the almost singular pattern of an
observable as a function of temperature.} temperature dependence of the most
relevant thermodynamical observables. Changes of state, like melting, 
condensation and so on are examples of first order phase transitions, the 
spontaneous magnetization of a paramagnetic material
when temperature is lowered below the so-called Curie temperature is an
example of a second order phase transition\footnote{According to the 
Ehrenfest's definitions, the first or the second derivative of the 
Helmoltz free energy with respect to temperature is discontinuous at a first
order transition point or at a second order transition point, respectively.}.
The usual framework of both theoretical and numerical 
investigations of phase transitions is mainly that of Gibbs' canonical 
ensemble (see Appendix A). 

Recently, the question of whether microscopic Hamiltonian dynamics displays 
some relevant change at a phase transition has been addressed by several works
(Antoni $\&$ Ruffo 1995; Caiani et al. 1997; Dellago $\&$ Posch 
1997, and earlier works therein cited; Gross 1997; Casetti et al. 2000; 
Cerruti-Sola, Clementi $\&$ Pettini 2000). 
The dynamical approach has its natural statistical mechanical counterpart 
in the microcanonical ensemble (see Appendix A). After a relaxation time which
depends on the initial condition, the time averages of observables  
computed along numerical phase space trajectories  provide good estimates 
of their microcanonical counterparts.
In the following Sections we adopt this dynamical approach to 
investigate if also in self-gravitating $N$-body
systems a phase transition is present and if
the dynamics shows a corresponding qualitative change.

\smallskip
\subsubsection{Caloric curve}
\smallskip
The first goal of our numerical computations was to obtain the caloric 
curves of different self-gravitating systems resulting 
from different choices of the
parameters. The caloric curve $T(E)$ (temperature {\it vs} energy) represents
a basic link between the microcanonical statistical ensemble and
thermodynamics, thus 
between dynamics and thermodynamics. From the entropy definition  
$S(E, V, N)= k_B \log \omega_N(E)$ in the microcanonical ensemble, $\omega_N$
being given by Eq.(\ref{micro}), temperature is derived as  
\begin{equation}
\frac{1}{T(E)}=\frac{\partial S(E)}{\partial E}~,
\label{temperature}
\end{equation}
which, for systems described by standard Hamiltonians 
$H=\sum_i \frac{1}{2}p_i^2 + U(q)$,  yields the  expression
\begin{equation}
 T=\left[\left(\frac{3N}{2}-1\right)\langle K^{-1}\rangle_\omega\right]^{-1}~,
\label{T_omega}
\end{equation}
where $\langle K^{-1}\rangle_\omega$ is the microcanonical average of the
inverse of the kinetic energy $K$.
The replacement of this average by means of the time average
$\langle K^{-1}\rangle_t$ provides the dynamical estimate of the temperature.

With the equivalent definition of microcanonical entropy as
$S(E, V, N)= k_B \log \Omega_N(E)$ (see Appendix A),  
Eq.(\ref{temperature}) yields the more familiar expression
\begin{equation}
 T=\frac{2}{3N}\langle K\rangle_{_\Omega}~,
\label{T_Omega}
\end{equation}
relating temperature with the mean kinetic energy per degree of freedom.
In our numerical computations, the two temperatures have been always found 
almost coincident, well
within the theoretically expected deviation of ${\cal O}(1/N)$.
Anyway, in the results reported here, $T$ is computed according to 
Eq.(\ref{T_omega}).

The Fourier-truncated model, due to the finite number of modes ${\cal N}_w$
considered, underestimates the potential energy with respect to the direct
model. However, since an additive constant in the Hamiltonian 
(\ref{Hbar}) does not
affect the dynamics, we can shift a-posteriori the energy scale by computing
the average energy difference of a set of random configurations whose
potential energy is computed according to both Eqs.(\ref{Hvera}) and 
(\ref{Hbar}).
Thus, in the following,
in order to compare by superposition the results of the two models, the values
of the energy densities obtained from
the Fourier truncated model have been shifted
towards the energy densities of the direct simulation.

\begin{figure}
\centerline{\psfig{file=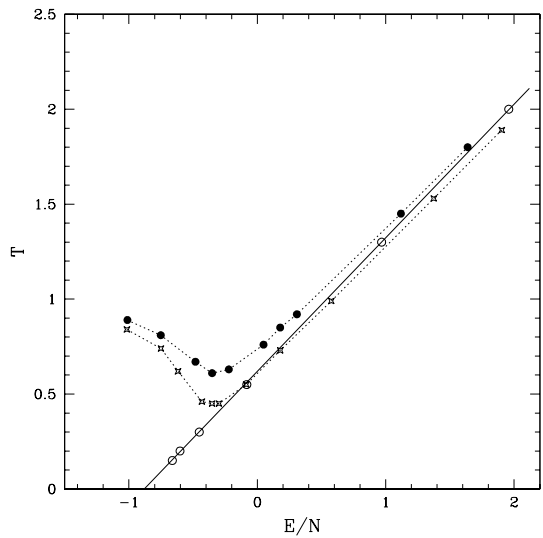,height=8cm,clip=true}}
\caption{ Caloric curve $T(E)$. $\varrho = 4\cdot 10^{-4}$ is the density
$N/L^3$. Here $N=50$ and ${\cal N}_w=7$. Starred squares refer to periodic 
boundary conditions, full circles refer to reflecting boundary conditions 
(box). Open circles represent MonteCarlo simulation results for the
Fourier-truncated model in the canonical
ensemble.  }
\label{fig_TE1}
\end{figure}

In Fig. \ref{fig_TE1} the caloric curve is reported for the Fourier-truncated
model (\ref{Hbar}) with ${\cal N}_w=7$, $N=50$ interacting objects and
$L=50$, so that $\varrho =4\cdot 10^{-4}$. The results obtained with periodic
and reflecting boundary conditions are synoptically displayed together with
the outcomes of MonteCarlo simulations in the case of reflecting boundary 
conditions. In the case of MonteCarlo simulations, temperature is the input
parameter and the average total energy is the outcome; anyway, since there is
a one-to-one correspondence between temperature and the average total energy,
the results can be reported without ambiguity also on the $T$ {\it vs} $E$
plane. The dynamical (microcanonical) caloric curves display the same 
qualitative features
and are even quantitatively not very different one from another. The low
energy branch has a negative slope, meaning that the specific heat is
negative. At some value of the energy per degree of freedom, the
slope of $T(E)$ becomes positive and the curve bends towards an asymptotic 
straight line of slope $2/3$, proper to a gas of independent particles.

\begin{figure}
\centerline{\psfig{file=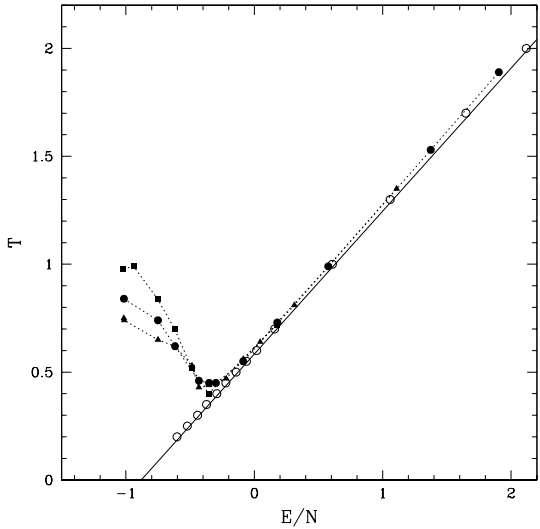,height=8cm,clip=true}}
\caption{Fourier-truncated model.
Comparison between the caloric curves obtained with  ${\cal N}_w=5$
(full triangles), ${\cal N}_w=7$ (full circles) and ${\cal N}_w=10$ 
(full squares) with periodic boundary conditions, at
$\varrho = 4\cdot 10^{-4}$ and $N=50$. Open circles are MonteCarlo results
with ${\cal N}_w=7$.    }
\label{fig_TE3}
\end{figure}

Fig. \ref{fig_TE3} shows the effect of different truncations in 
Eq. (\ref{Hbar}): qualitatively, the results are satisfactorily stable,
quantitatively, the larger ${\cal N}_w$, the closer is $dT/d\epsilon$ (with
$\epsilon =E/N$) to the newtonian slope $-2/3$.

\begin{figure}
\centerline{\psfig{file=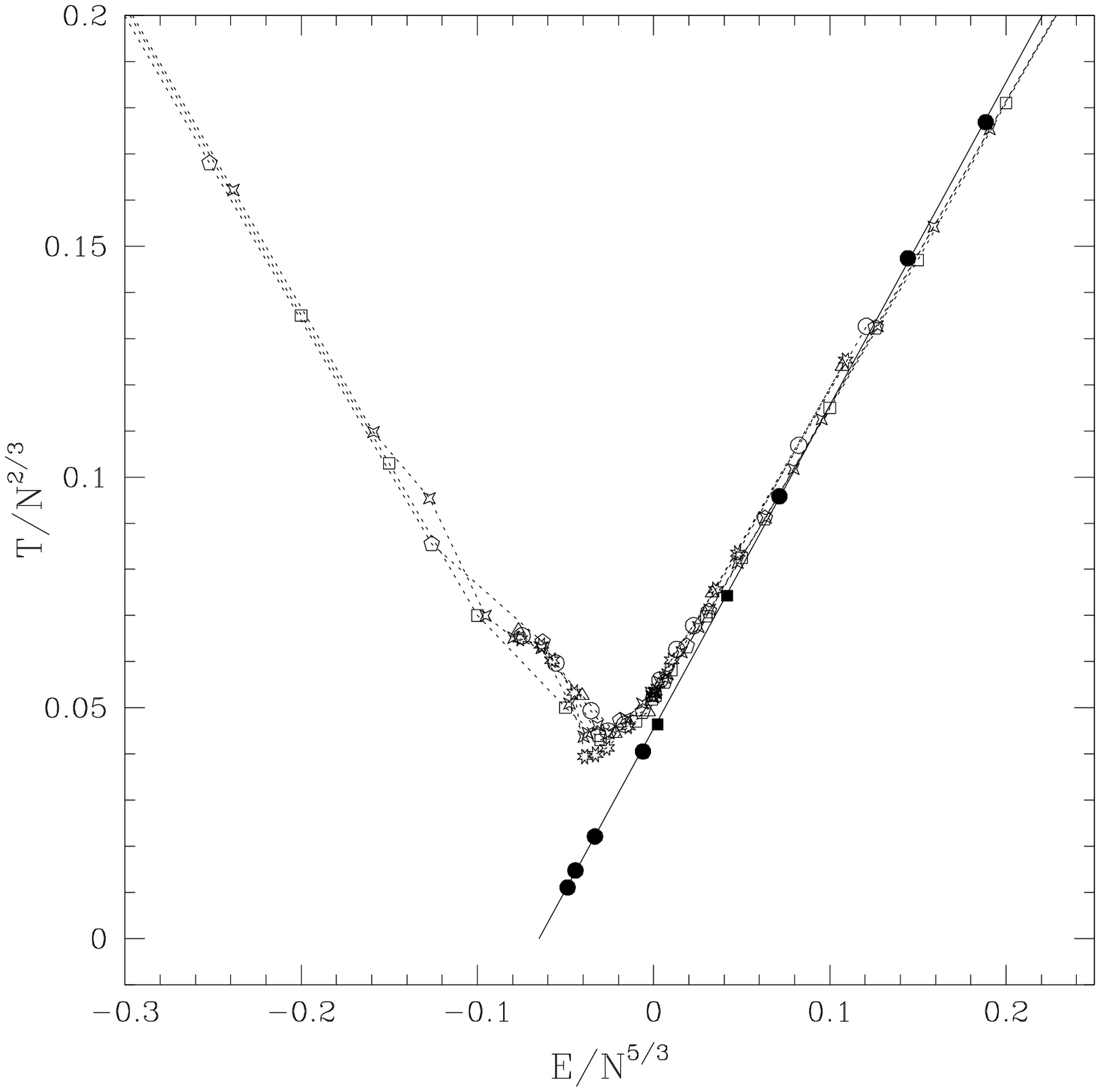,height=8cm,clip=true}}
\caption{Comparison between the results of the Fourier-truncated model and
the direct simulation with standard softening.
The results of the former have been suitably shifted (see text).
 Here $\varrho = 4\cdot 10^{-4}$
and the boundary conditions are reflecting. Fourier-truncated model with  
${\cal N}_w=7$: $N=25$ (open triangles), $N=50$ (open circles), $N=100$ 
(starred polygons). Direct $N$-body simulation with softening parameter 
$\eta =0.01$: $N=500$ (starred squares), 
$N=1000$ (open squares), $N=2000$ (open pentagons). MonteCarlo
canonical results for the Fourier-truncated model: $N=50$ (full circles),
 $N=100$ (full squares).  }
\label{fig_TE4}
\end{figure}

\begin{figure}
\centerline{\psfig{file=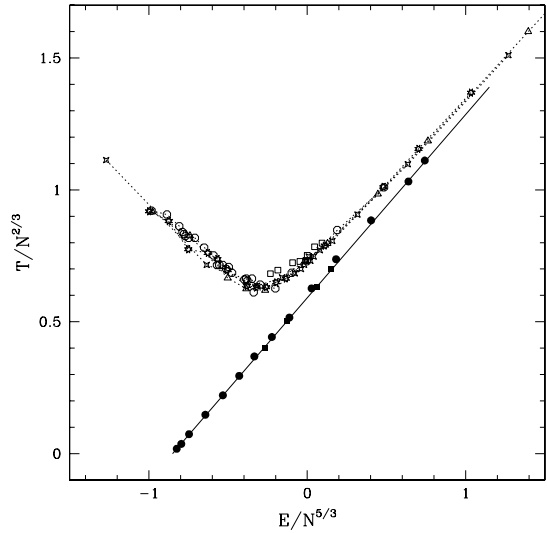,height=8cm,clip=true}}
\caption{Comparison between the results of the Fourier-truncated model and
the direct simulation with standard softening.
The results of the former have been suitably shifted (see text).
 Here $\varrho = 1$
and the boundary conditions are reflecting. Fourier-truncated model with
${\cal N}_w=7$: $N=25$ (open triangles), $N=50$ (open circles), $N=100$
(starred polygons). Direct $N$-body simulation with softening parameter
$\eta =0.01$: $N=100$ (open squares), $N=500$ (starred squares). 
MonteCarlo
canonical results for the Fourier-truncated model: $N=50$ (full circles),
 $N=100$ (full squares).  }
\label{fig_TE5}
\end{figure}

Notice that Figs. \ref{fig_TE4} and \ref{fig_TE5} display also a slope
$-2/3$ of the negative-$c_V$ branches, in agreement with the prediction of
the virial theorem, in the regime in which the effect of the box is negligible
(large negative energy values). 

Fig. \ref{fig_TE4} and Fig. \ref{fig_TE5} summarize a lot of information.
They show the caloric curves obtained for different values of $N$, for the
newtonian $N$-body simulations and for the Fourier-truncated model computed 
with the canonical MonteCarlo algorithm and with the microcanonical 
simulations.
The data collapsing\footnote{This is standard jargon in the numerical study 
of phase transitions, and it means that when different sets
of results are compared after a suitable rescaling with the parameter that
varies from a set to another, then all the results crowd on the same curve.}
is obtained by rescaling the energy with $N^{5/3}$ and temperature with
$N^{2/3}$ (notice that temperature is already an energy divided by $N$).
The simple argument given in Section \ref{Nscaling} appears correct.

There is a very good agreement between the results obtained with the 
Fourier-truncated model and those obtained with the direct $N$-body 
simulations. This is a very interesting result because of the relatively small
number of Fourier modes considered, and in view of adopting the 
Fourier-truncated model for the analytic computation of statistical mechanical
averages.

It is remarkable  that the change of sign of the slope of the caloric curve
implies the existence of a phase transition. In fact, $T(E)$ can only change
the sign of its slope either through a ``V''-shaped pattern or through an
``U''-shaped pattern: in the former case, the derivative
$\partial T/\partial E$ is discontinuous at the cuspy point so that the 
specific heat makes a discontinuous jump (first-order phase transition), 
whereas in the latter case the same derivative vanishes at the minimum so that
the specific heat diverges (second order phase transition).

The comparison among the dynamical (microcanonical) and the MonteCarlo 
(canonical) caloric curves shows a remarkable fact: the canonical caloric
curve seems indistinguishable from that of a perfect gas, no reminiscence in 
the canonical results seems to be left of the feature shown by the 
microcanonical curves. As already mentioned throughout this paper, a negative
specific heat is strictly forbidden in the canonical ensemble; however, 
in the case of a two dimensional model, it has been observed (Lynden-Bell $\&$
Lynden-Bell 1977) that the appearance of negative specific heat in the
microcanonical ensemble corresponds to a signal of a phase transition in the
canonical ensemble. Nothing similar is found here. This is a remarkable result:
though a-priori the ensemble inequivalence was expected, such a radical loss
of any signal of the transition in the canonical ensemble was unexpected. 
\smallskip
\subsubsection{Specific heat}
\label{spec-heat}
\smallskip
From the above given definition of temperature the microcanonical specific
heat is computed according to the formula
\[
\frac{1}{C_V}=\frac{\partial T(E)}{\partial E}~.
\] 
Hence the knowledge of the caloric curve implies the knowledge 
also of the specific heat, but in practice the points on the $T - E$ plane 
are inevitably affected by a ``noise'' which can make too rough the computation
of $C_V$
as the numerical derivative of the caloric curve. The interest of an 
independent numerical derivation of the specific heat is therefore twofold:
on one side, it constitutes a necessary and useful check of the previous 
results, on the other side it should give more information about the order of
the phase transition.

The already mentioned Laplace transform technique (Pearson et al. 1985)
yields the following expression 
\begin{eqnarray}
c_V&=&\left(\frac{C_V}{N}\right)_\omega \label{cvmicro} \\
&=&\left(\frac{3}{2}N - 1\right)\left[N \left(\frac{3}{2}N - 1
\right) - N \left(\frac{3}{2}N - 2\right)\frac{\langle K^{-2}\rangle}{\langle 
K^{-1}\rangle^2}\right]^{-1}\nonumber
\end{eqnarray}
which has been used in our numerical simulations; $K$ is the total kinetic
energy.

The  specific heat {\it vs} energy qualitatively displays a pattern 
in full agreement with what can be guessed from the caloric 
curves. At small energy values a branch of negative specific heat is found.
At an energy value slightly below the minimum of the  
caloric curve, the specific heat shows two marked spikes, stressing the 
occurrence of a sudden jump between negative and positive
values. The high energy asymptotic value is $3/2$, in quantitative agreement
with the high energy slope of $2/3$ of the caloric curve.
Again the data collapsing is obtained by scaling the total energy with 
$N^{5/3}$. 

\begin{figure}
\centerline{\psfig{file=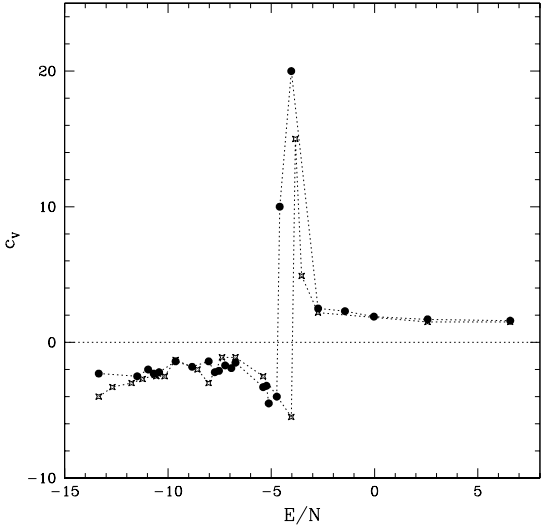,height=8cm,clip=true}}
\caption{Fourier-truncated model. Specific heat  at $N=50$, 
$\varrho = 1$ and ${\cal N}_w=7$. Results obtained with reflecting boundary 
conditions (full circles) are compared with those obtained with periodic 
boundary conditions (starred squares).  }
\label{fig_cv1}
\end{figure}

In Fig. \ref{fig_cv1} the specific heat obtained with periodic 
and reflecting boundary conditions respectively is reported {\it vs} energy 
density for the Fourier-truncated model. Also through this observable it is
confirmed that there is a weak sensitivity of the outcomes upon the boundary
conditions, at least as far as thermodynamical observables are concerned. 

\begin{figure}
\centerline{\psfig{file=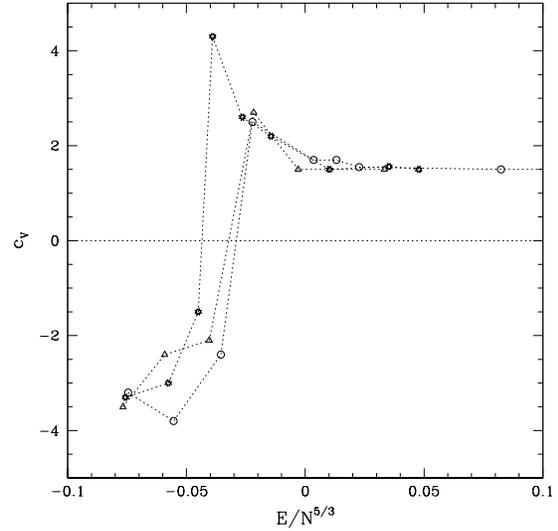,height=8cm,clip=true}}
\caption{Fourier-truncated model. Specific heat obtained with  
$\varrho = 4\cdot 10^{-4}$ and ${\cal N}_w=7$. Reflecting boundary conditions.
Comparison among: $N=25$ (open triangles), $N=50$ (open circles) and $N=100$
(starred polygons).    }
\label{fig_cv2}
\end{figure}

\begin{figure}
\centerline{\psfig{file=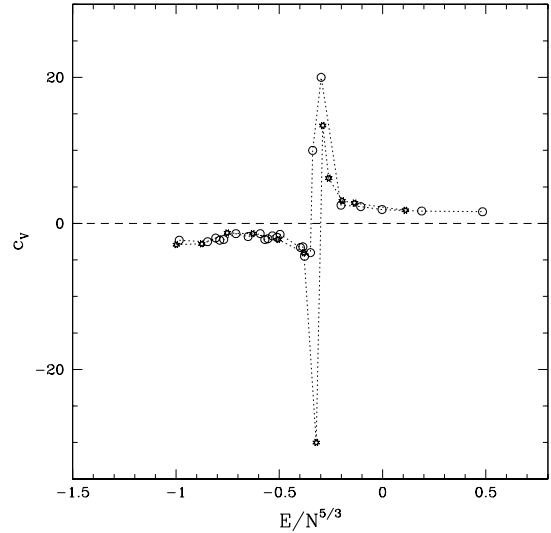,height=8cm,clip=true}}
\caption{Fourier-truncated model. Specific heat obtained at  
$\varrho = 1$, ${\cal N}_w=7$ and with reflecting boundary 
conditions. Open circles refer to $N=50$ and starred polygons to $N=100$. }
\label{fig_cv3}
\end{figure}

\begin{figure}
\centerline{\psfig{file=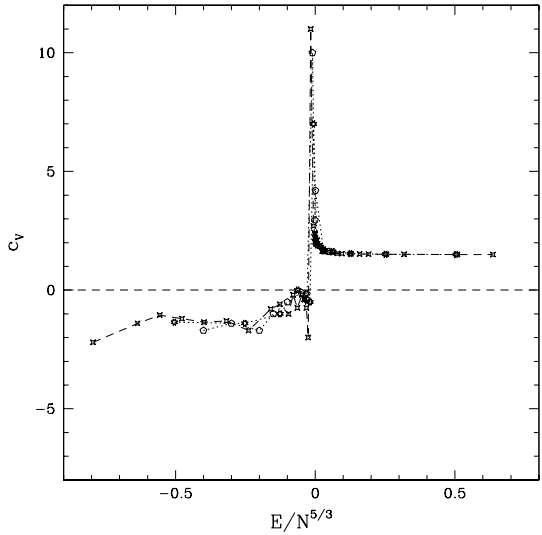,height=8cm,clip=true}}
\caption{Direct $N$-body simulations. Reflecting boundary conditions and
$\varrho = 4\cdot 10^{-4}$ and $\eta=0.01$. 
Comparison of the specific heat obtained at
$N=500$ (starred squares), $N=1000$ (pentagons) and $N=2000$ (starred 
polygons). }
\label{fig_cv2p}
\end{figure}

\begin{figure}
\centerline{\psfig{file=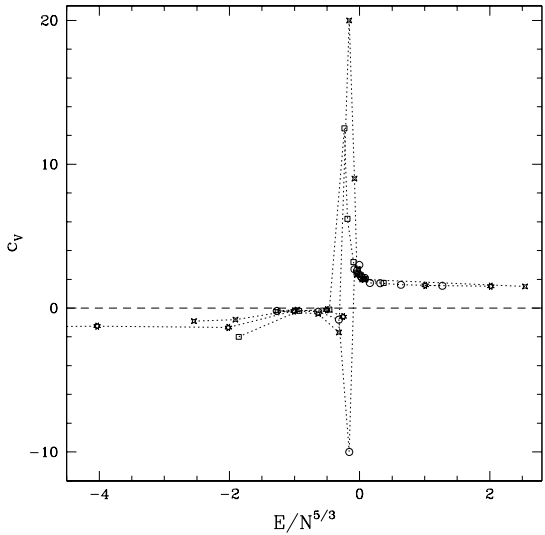,height=8cm,clip=true}}
\caption{Direct $N$-body simulations. Reflecting boundary conditions,
$\varrho = 1$. Comparison of the specific heat obtained at
$N=100$ and  $\eta=0.01$ (open squares), $N=500$ and $\eta=0.01$ (open 
circles), $N=500$ and $\eta=0.02$ (starred squares), and 
$N=2000$ and $\eta=0.02$ (starred polygons). Here $\eta$ is varied as a 
stability check on the results. }
\label{fig_cv3p}
\end{figure}

For the Fourier-truncated model, Fig. \ref{fig_cv2} and Fig. \ref{fig_cv3} 
show -- at different values of the density -- how the pattern of $c_V$ 
changes when $N$ is increased. 
The corresponding results obtained for the direct $N$-body simulations are 
reported in Fig. \ref{fig_cv2p} and Fig. \ref{fig_cv3p}.
A common feature of these results is the ambiguous behaviour of the
height of the peaks of $c_V$, close to the transition point, when $N$ is
increased, at variance with other systems with short-range interactions
(Caiani et al. 1998; Cerruti-Sola et al. 2000).
The doubtful absence of a systematic increase with $N$ of the peak of $c_V$
could be due to three main reasons: {\it i)} a very fine tuning of the energy
value could be necessary and the right energy values could have been missed;
{\it ii)}  the transition could be too mild, for example with only a 
logarithmic divergence of $c_V$; {\it iii)}  the pattern of $T(E)$ could be
``V''-shaped, thus $\partial^2S/\partial E^2$ should make a finite jump at 
the transition point, with the physical consequence that the system should 
undergo a first order phase transition.

\begin{figure}
\centerline{\psfig{file=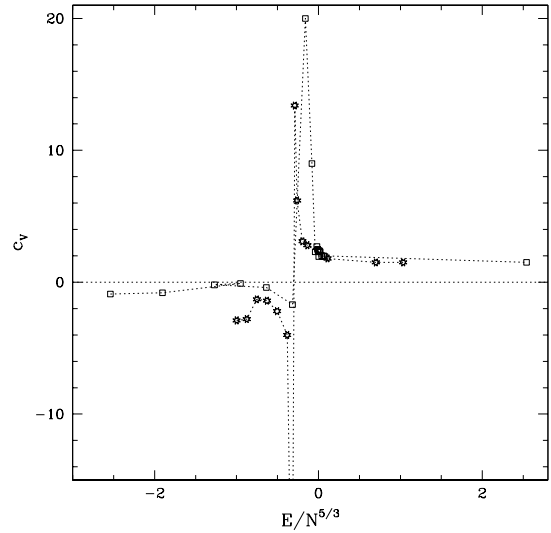,height=8cm,clip=true}}
\caption{Comparison of the specific heat obtained at   
$\varrho = 1$ and with reflecting boundary conditions:
 Fourier-truncated model with $N=100$ (starred polygons), direct $N$-body 
simulations with $N=100$ (open triangles) and with $N=500$ (open squares). }
\label{fig_cv-comp2}
\end{figure}
The high energy branches of $c_V$, above the transition energy, obtained 
with the 
Fourier-truncated model and with the direct $N$-body simulation are in very
good agreement, whereas the quantitative agreement is less good in the branch
of negative values. Notice that in the Fourier-truncated model there is a 
lower bound to the energy values which depends on the number of modes retained 
in the truncated Fourier expansion, the larger the number of modes, the lower 
is the minimum attainable value of negative energies\footnote{Also for a  
softened newtonian system there is a lower bound on the total energy.
With the values of the softening chosen here, the lower bound is
much smaller than that of the Fourier-truncated model.}. 
Figures \ref{fig_cv1} through \ref{fig_cv2p} show that the results of the 
computations of $c_V$ according to Eq.(\ref{cvmicro}) are in very good
agreement with the corresponding patterns of $T(E)$. In both the direct 
newtonian simulations and in the Fourier-truncated model, 
a kink in the negative slope region of $T(E)$ is found and is 
confirmed by the $c_V(E/N^{5/3})$ pattern close to the transition point.
We attribute the appearance of this feature to the competing effects of the 
fixed size of the box and the increasing (with energy) gravitational radius of
the system.
  
\smallskip
\subsubsection{Order parameter}
\smallskip
Both in the theoretical and numerical studies of phase transitions, the choice
of a good order parameter for a given system is a main point. An order 
parameter is typically a collective variable that bifurcates at the transition
temperature $T_c$. A classic example is the spontaneous magnetization of a
ferromagnetic material: it is zero above $T_c$, then it suddenly bifurcates 
away from zero at $T_c$ and increases as temperature is lowered below $T_c$.
True singularities exist only in the $N\rightarrow\infty$ limit; at finite $N$
they are smoothed but it is still possible to detect the existence of a phase 
transition by changing $N$ and observing if the smoothed signals tend to 
sharpen or not.

In many cases the definition of an order parameter is not trivial. A-priori 
this is true also for self-gravitating $N$-body systems. However,
let us notice an interesting property of the Hamiltonian (\ref{Hbar}) and of
the equations of motion 
(\ref{eqmotion}): at large $N$ the coefficients $S_{l,m,n}^{(i)}$ can be
approximated by the $S_{l,m,n}$ with an error ${\cal O}(1/N)$, and  
the replacement of the $S_{l,m,n}$ by the averages $\langle S_{l,m,n}\rangle$, 
computed through a consistency equation (see Appendix B), {\it decouples} all 
the degrees of freedom. The Hamiltonian (\ref{Hbar}) is approximated by
\[
 H_{mf} 
= \sum_{i=1}^N\frac{1}{2}\left( p_{xi}^2 + p_{yi}^2 + p_{zi}^2\right)
\]
\[
- \frac{16}{\pi L}\sum_{i=1}^N\sum_{l,m,n=1}
^{{\cal N}_w} \frac{\langle S_{l,m,n}\rangle}{ (l^2+m^2+n^2)}
\sin ( k_l x_i)\sin (k_m y_i)\sin (k_n z_i)~.
\]
\begin{equation}
\label{Hmf}
\end{equation}
which is now in the form of a sum over independent degrees of freedom, because
the $\langle S_{l,m,n}\rangle$ are collective variables or, equivalently, 
order parameters, that take the same values for all the coordinates 
$x_i,y_i,z_i$. 
The replacement of the coefficients $S_{l,m,n}$ with the averages 
$\langle S_{l,m,n}\rangle$ in Hamiltonian (\ref{Hmf}) is a typical mean-field
approximation in  statistical mechanics (the $\langle S_{l,m,n}\rangle$ 
are like Weiss molecular field of the early times of statistical mechanics 
of magnetic materials). 
The recasting of Hamiltonian (\ref{Hvera})
into the approximate form (\ref{Hmf}), which can be as precise as we may
want if $N$, ${\cal N}_w$ and $L$ are sufficiently large, is of great
prospective interest for theoretical -- analytic or semi-analytic -- 
computations of statistical mechanical kind (see Appendix B).

Out of the huge family of order parameters  $\langle S_{l,m,n}\rangle$,  
it is natural to define the following two global order parameters in analogy
with other more standard contexts (Caiani et al. 1998; Cerruti-Sola et al. 
2000)
\begin{equation}
M_1=\frac{1}{N}
\sum_{l,m,n=1}^{{\cal N}_w}\frac{l+m+n}{ (l^2+m^2+n^2)}\ 
\langle S_{l,m,n}\rangle
\label{M1}
\end{equation}
and            
\begin{equation}
M_2=\frac{1}{N}
\sum_{l,m,n=1}^{{\cal N}_w}\frac{l+m+n}{ (l^2+m^2+n^2)}\ \langle\vert 
 S_{l,m,n}\vert\rangle~.
\label{M2}
\end{equation}
The coefficients $(l+m+n)/(l^2+m^2+n^2)$ are the sum of the
corresponding coefficients in the equations of motion (\ref{eqmotion}) for 
the three variables $x,y,z$. They decrease as the norm  of the 
wavevector ${\vec k}=(l,m,n)$ increases, compensating  the growing 
number of the $S_{l,m,n}$ terms as the norm of ${\vec k}=(l,m,n)$ increases.
We easily realize that $M_1$ cannot be very sensitive to a change
of the particle distribution in the box because the dishomogeneities measured
at different scales tend to cancel each other. Thus, to get rid of this
problem, $M_2$ is defined through the sum of the absolute values of the 
$S_{l,m,n}$. Actually, the numerical computations, performed at different 
energies and $N$, confirmed that $M_1$
is always very close to its minimum value (which depends on ${\cal N}_w$).
At variance, the order parameter $M_2$ is very effective to detect the 
clustering transition and to make strict the analogy between this transition
and a thermodynamic phase transition.

In order to measure the deviation from the homogeneous spatial 
distribution of particles, it is convenient to slightly modify the order 
parameter as follows: we denote with $M_0$ the value of $M_2$ 
that corresponds to a uniform density of particles in the box, then
we modify the order parameter $M_2$ to $\mu_2 = {\nu}(M_2 - M_0)$, where
${\nu}$ is a normalization constant. Thus $\mu_2$ varies in
the interval $[0,1]$, being zero for a uniform distribution of particles and
one for a maximally clustered configuration. Such a normalization makes easy
the comparison of the results obtained at different $N$.
In the case of the Fourier truncated model, the maximally clustered 
configuration is obtained when all the particles are concentrated in a cell
whose side is equal to the shortest wavelength. However, for simplicity, 
we empirically normalized $\mu_2$ by using the largest value of $M_2$ measured
at the lowest accessible energy for the given truncation of the Fourier 
expansion.

The values $M_0$ of $M_2$, corresponding to a uniform density of 
particles in the box, have been numerically computed by averaging $M_2$ on
a few hundreds of uniformly distributed random
configurations of $50, 100, 200$ and $400$ particles.
For $N=50, 100, 200, 400$ and ${\cal N}_w=7$ we found 
$M_0=3.49, 2.60, 1.97, 1.54$ respectively. 

The averages  $\langle\vert S_{l,m,n}\vert\rangle$ are numerically computed 
as time averages to give $M_2$, and the order parameter $\mu_2$ is worked out 
as a function of the energy and of the number of particles by using the above 
reported values of $M_0$.

The same order parameter $\mu_2$ has been computed for the direct $N$-body
simulations. Also in this case we considered ${\cal N}_w=7$, though in
principle this choice is arbitrary because ${\cal N}_w$ is independent of 
the dynamics. Since a phase transition is a collective effect, driven by
the long wavelength modes, there is no need for a large value of ${\cal N}_w$.

\begin{figure}
\centerline{\psfig{file=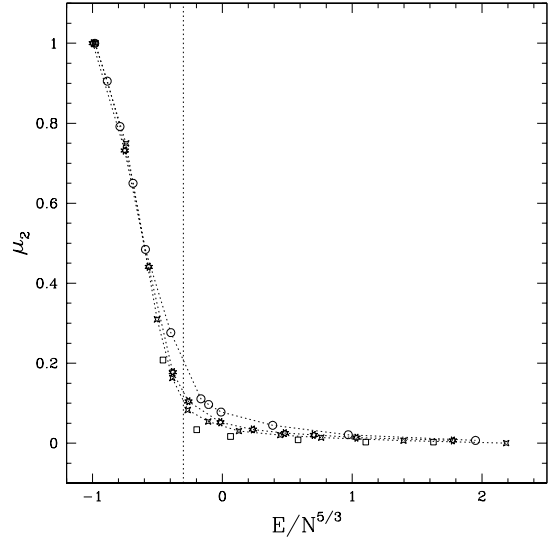,height=8cm,clip=true}}
\caption{  Order parameter $\mu_2$ for the 
Fourier-truncated model.  
$\varrho = 1$, ${\cal N}_w=7$ and reflecting boundary 
conditions. Open circles refer to $N=50$, starred polygons to $N=100$, 
starred squares to $N=200$ and open squares to $N=400$.    }
\label{fig_M2}
\end{figure}

\begin{figure}
\centerline{\psfig{file=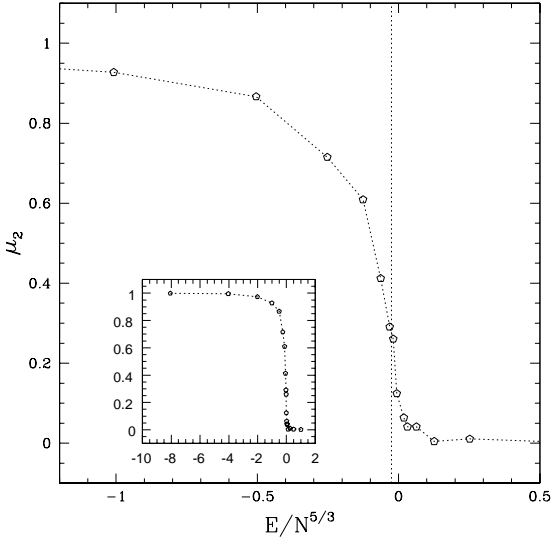,height=8cm,clip=true}}
\caption{ Order parameter $\mu_2$ for the 
direct $N$-body simulation.  
$\varrho = 4\cdot 10^{-4}$, ${\cal N}_w=7$ and reflecting boundary 
conditions. Here $N=2000$. }
\label{fig_M2p}
\end{figure}

The final results are reported versus $E/N^{5/3}$ in Fig. \ref{fig_M2} for 
the Fourier-truncated model, and in Fig. \ref{fig_M2p} for the direct 
$N$-body simulations.

The data reported in these figures display typical patterns
that are found in presence of a phase transition in other contexts.  
In fact, the order parameter $\mu_2$ displays a classic bifurcation pattern 
at the transition point, smoothed by the finite and not very large number of 
particles but with a clear tendency to become sharper at larger $N$ as is 
shown  by Fig. \ref{fig_M2}. The simulation of the direct $N$-body system
allows to work out the bifurcation pattern of the order parameter in a larger
energy interval, because there is no lower bound to the energy\footnote{
As already noticed, the smoothing parameter entails a lower bound to the 
energy also in this case, but it occurs at a very large negative value.}. 
The inset of Fig. \ref{fig_M2p} displays the pattern of $\mu_2$ on a large 
energy interval.
The vertical dotted lines in both figures indicate the energies of
the peaks of specific heat. In general, the inflection point of the order
parameter and the location of the peak of the specific heat do not coincide
but tend to get closer at larger $N$, thus making more precise the phase
transition point.
The bifurcation pattern of $\mu_2$ gives a reliable indication about the 
order of the phase transition: it strongly indicates a second order 
transition, thus eliminating the ambiguity of the information given by $c_V$.
In fact, in presence of a first order transition, the order parameter $\mu_2$
should make a finite jump at the transition point which is clearly 
not our case. We have formulated in Section 
\ref{spec-heat} two possible explanations ({\it i)} and {\it ii)}) -- 
alternative to the first order phase transition -- of the behaviour of $c_V$ 
which can reconcile the outcomes on $c_V$ with those on $\mu_2$. 

Notice that $M_2$ is also a measure of the average modulus 
$\langle\vert{\vec F}_i\vert\rangle$ of the force acting
on each mass and -- independently of $N$ -- this is never negligible for a
bounded and clustered system, whereas in the homogeneous phase
$\langle\vert{\vec F}_i\vert\rangle$ depends on the relative density 
fluctuations and thus it is expected to decrease with increasing $N$.
The physical consequence is that, in the homogeneous phase, the larger $N$ the
smaller is the relative average weight of the potential energy with respect 
to the kinetic energy so that the system behaves like a collection of almost  
independent particles, i.e. not far from a perfect gas. 

\subsection{Equation of state}
\label{pv}
The equation of state of a system contains the basic information of how
pressure changes with volume at constant temperature. This information is
contained in a family of isothermal curves in the $P-V$ plane. If a phase
transition is there, then the isotherms at $T<T_c$ display a substantial
difference from those at $T>T_c$, as an example, the isotherms of the van 
der Waals equation, describing the liquid-vapour transition, display a kink
at $T<T_c$ (Huang 1963).

Since dynamics has its natural statistical counterpart in the microcanonical 
ensemble, where energy rather than temperature is the independent variable,
the isotherms are replaced by isoenergetic curves in the $P-V$ plane.

From thermodynamics we have 
\[
\left(\frac{\partial S}{\partial V}\right)_E=\frac{P}{T}
\]
and using the above given definition of temperature one gets
\[
k_B T = \frac{\Omega_N}{\omega_N}~,
\]
where $\Omega_N$ is defined in the Appendix A. Hence
\[
P=\left(\frac{1}{\omega_N}\frac{\partial \Omega_N}{\partial V}\right)_E
\]
whence, with the aid of the already mentioned Laplace transform technique, 
one obtains (Pearson et al. 1985)
\begin{equation}
(P)_\omega = \frac{N}{V}(k_B T)_\omega - \left\langle \left(\frac{\partial U}
{\partial V}\right)_E K^{-1}\right\rangle_\omega 
\langle K^{-1}\rangle_\omega^{-1}~,
\label{eq_stato}
\end{equation}
which is the microcanonical equation of state yielding pressure {\it vs} 
volume at constant energy; $K$ is the total kinetic energy and $U$ is the 
potential energy. 
 Temperature is given by Eq.(\ref{temperature}) as a function
of the energy. Again, the averages in Eq. (\ref{eq_stato}) are computed 
as time averages along the numerical phase space trajectories.

\begin{figure}
\centerline{\psfig{file=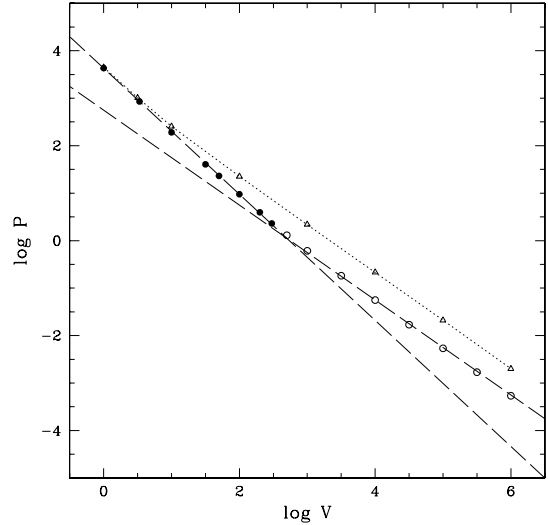,height=8cm,clip=true}}
\caption{ Equation of state: {\it pressure} vs. {\it volume}.
Fourier-truncated model with $N=100$, ${\cal N}_w=7$ and reflecting boundary 
conditions.  Open triangles refer to ${\tilde E}=3174.8$,
full and open circles to ${\tilde E}=793.7$. $\tilde E$ are not shifted
(see text). 
Full circles denote points that correspond
to negative specific heat, conversely open circles denote positive specific 
heat. Reference lines (long dashed) correspond to a polytropic $V^{-4/3}$ 
law and to a perfect gas $V^{-1}$ law respectively.}
\label{fig_PV}
\end{figure}

\begin{figure}
\centerline{\psfig{file=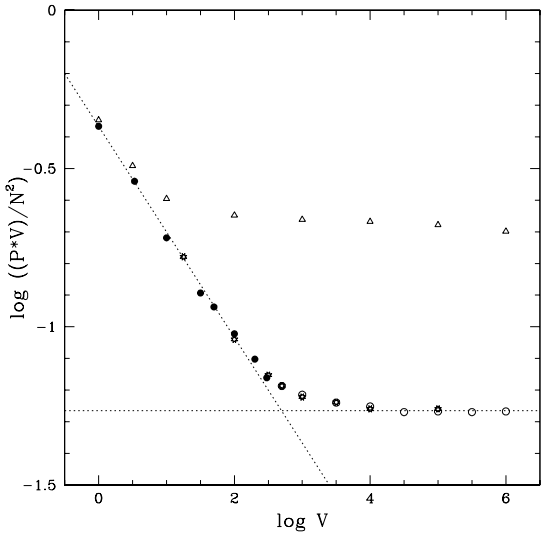,height=8cm,clip=true}}
\caption{Equation of state for the
Fourier-truncated model with $N=100$, ${\cal N}_w=7$ and reflecting boundary 
conditions. $PV/N^2$ is plotted vs. $V$ to put in evidence the transition from
a polytropic law to a perfect gas law. Symbols are the same as in Fig.
\protect\ref{fig_PV}. Starred polygons refer to $N=200$ and 
${\tilde E}=3174.8$.}
\label{fig_PxVV}
\end{figure}

By replacing $L=V^{1/3}$ in the potential part of the Hamiltonian 
(\ref{Hbar}),
and noting that after the
replacements $x_i=L\chi_i$, $y_i=L\eta_i$ and $z_i=L\zeta_i$ (where $\chi_i,
\eta_i, \zeta_i$ are adimensional variables) the arguments of the sines do 
not depend on $L$, the following simple result is found
\begin{equation}
\frac{\partial U}{\partial V} = -\frac{U}{3V}~,
\label{dUdV}
\end{equation}
so that the microcanonical equation of state finally reads
\begin{equation}
(P)_\omega = \frac{N}{V}(k_B T)_\omega + \frac{1}{3V} \left\langle 
\frac{U}{K}\right\rangle_\omega \langle K^{-1}\rangle_\omega^{-1}~.
\label{eq_state}
\end{equation}
Its explicit analytic computation seems hard but still feasible with the
aid of the mean-field decoupling of the degrees of freedom in Eq.(\ref{Hmf}),
however such an attempt is beyond our present aims and again we use dynamics
in order to estimate the microcanonical averages through time averages.
Moreover, by extracting out of $U$ its prefactor $V^{-1/3}$ 
[see Eq.(\ref{Hbar})], we see that Eq.(\ref{eq_state}) is of the form
\begin{equation}
P V = N k_B T - {\cal U}(E) V^{-1/3}~,
\label{Bonnor}
\end{equation}
where ${\cal U}(E)$ would be the outcome of the analytic computation of the
microcanonical averages, i.e. 
${\cal U}(E)=\frac{1}{3}\langle V^{1/3}U/K\rangle\langle K^{-1}\rangle$. 
Equation (\ref{Bonnor}) agrees with the already
proposed modification of Boyle's perfect gas law for self-gravitating systems
(see the next Section) which is here qualitatively rederived in the 
microcanonical ensemble and quantitatively computed numerically. The
results are reported in Fig.\ref{fig_PV}, where we have differently marked the
points that correspond to positive and negative specific heat respectively.
The one-to-one correspondence polytrope-negative specific heat (perfect gas
law-positive specific heat) is well evident.
The transition from a polytropic
$V^{-4/3}$ law to a $V^{-1}$ perfect gas law is well evident. This reflects 
the competition between the two terms in the r.h.s. of Eqs.(\ref{eq_state})
and (\ref{Bonnor}). 
Figure \ref{fig_PxVV} shows  $PV/N^2$ 
{\it vs} $V$, obtained at $N=100, 200$ and at different values of the energy 
$\tilde E$ which have not been shifted towards their corresponding newtonian 
values as 
it was done in the preceding Sections (the reason should be clarified by the 
concluding remarks of the present Section). In order to compare the results 
obtained at different $N$, $\tilde E$ is scaled with $N^2$ as well as $PV$.
The results of Fig. \ref{fig_PxVV} directly compare to Eq.(\ref{Bonnor}) 
and make clearer the cross-over between $V^{-4/3}$ and $V^{-1}$. 
Such a cross-over
seems rather smooth, though a discontinuity in the derivative $dP/dV$ cannot
be excluded, and this seems to confirm that the gravitational clustering 
phenomenon is a peculiar phase transition with respect to all the known 
laboratory phase transitions. 

\begin{figure}
\centerline{\psfig{file=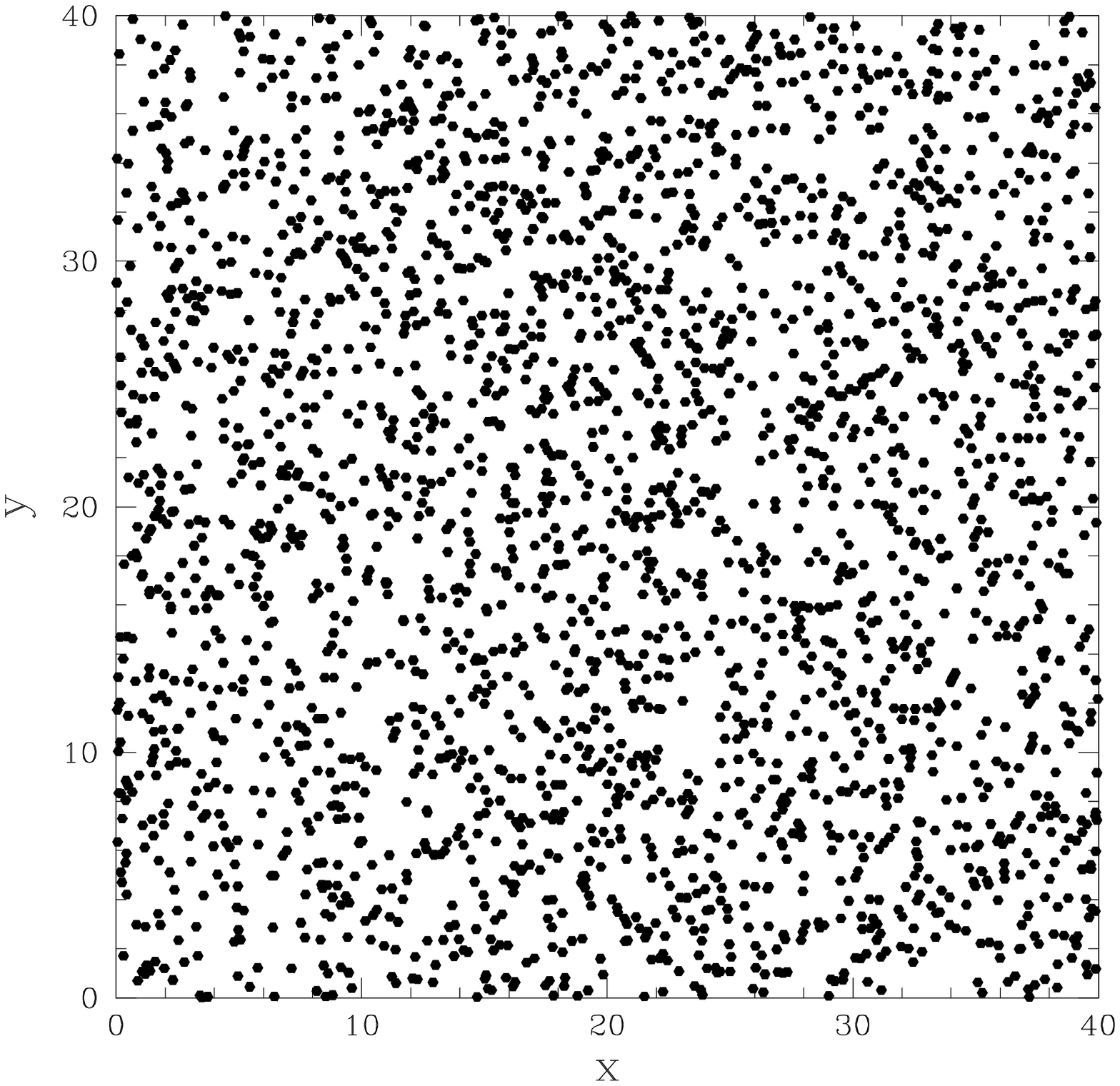,height=8cm,clip=true}}
\caption{Superposed snapshots of particle positions projected onto the 
$x-y$ plane for the 
Fourier-truncated model with $N=100$, ${\cal N}_w=7$ and reflecting boundary
conditions. Energy $E=793.7$ and $\log V=5.0$. Homogeneous phase (see Fig.
\protect\ref{fig_PxVV}).}
\label{x-y-hom}
\end{figure}

\begin{figure}
\centerline{\psfig{file=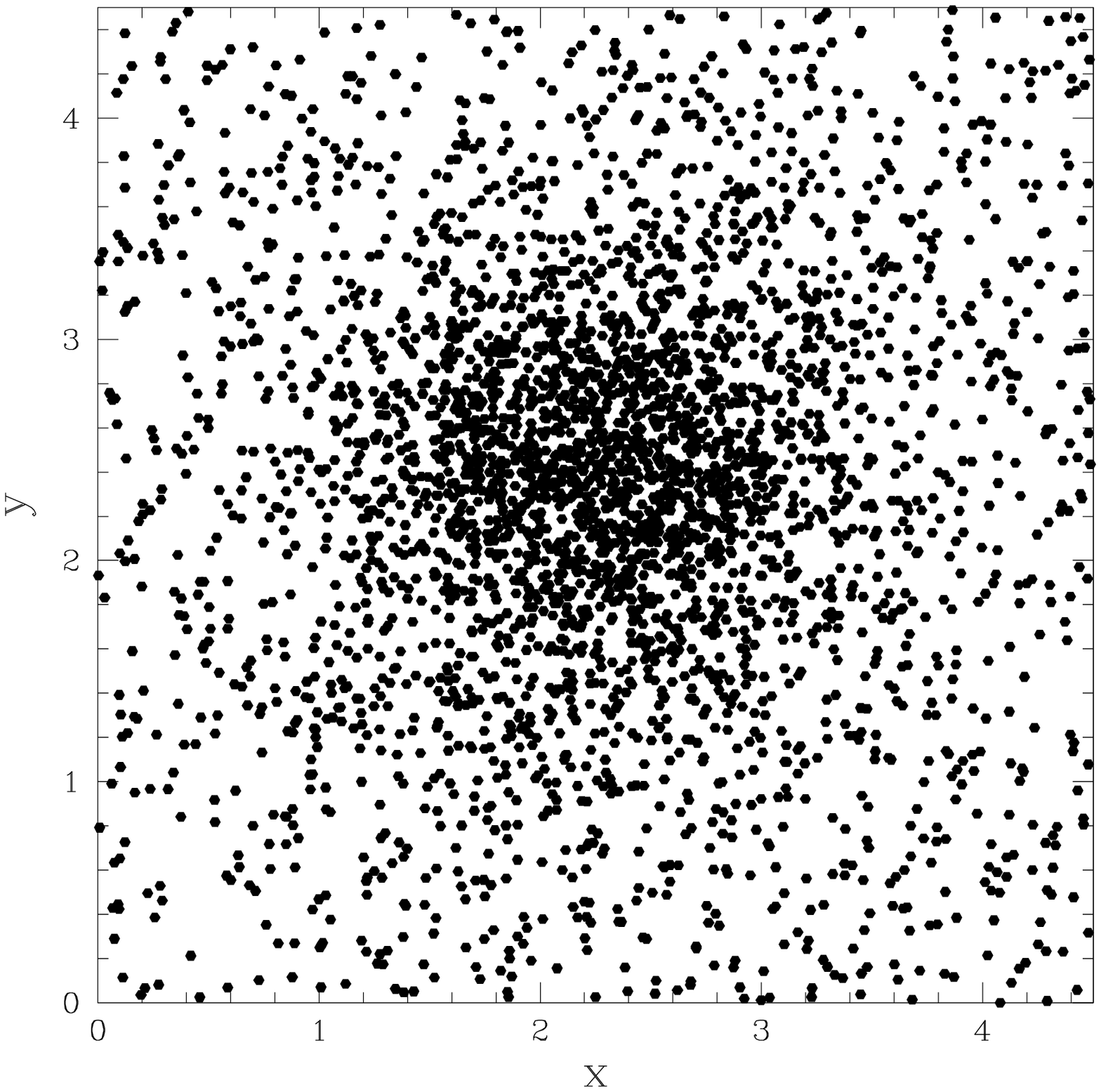,height=8cm,clip=true}}
\caption{Superposed snapshots of particle positions projected onto the 
$x-y$ plane for the 
Fourier-truncated model with $N=100$, ${\cal N}_w=7$ and reflecting boundary
conditions. Energy $E=793.7$ and $\log V=2.0$. Clustered phase (see Fig.
\protect\ref{fig_PxVV}).}
\label{x-y-clust}
\end{figure}

\begin{figure}
\centerline{\psfig{file=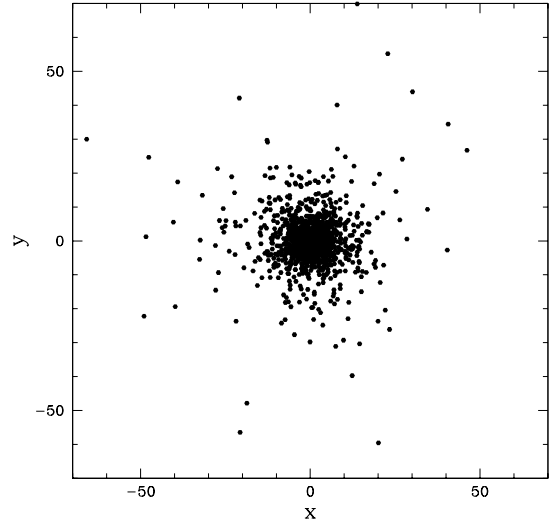,height=8cm,clip=true}}
\caption{Snapshot of particle positions projected onto the 
$x-y$ plane for the direct $N$-body system with $N=2000$ and reflecting boundary
conditions. $\varrho =4\cdot 10^{-4}$, $E/N^{5/3}=-0.5$ and $\log V=6.7$. 
Clustered phase (see Fig.\protect\ref{fig_TE4}).}
\label{x-y-core}
\end{figure}

\begin{figure}
\centerline{\psfig{file=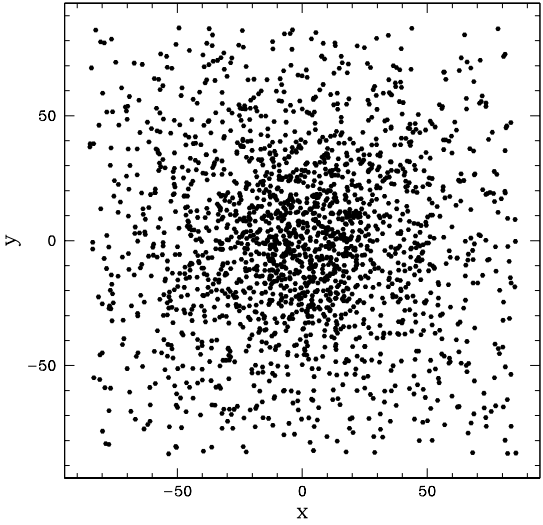,height=8cm,clip=true}}
\caption{Same as Fig.\protect\ref{x-y-core} but with $E/N^{5/3}=-0.03$.
Transition regime (see Fig.\protect\ref{fig_TE4}).}
\label{x-y-tran}
\end{figure}

\begin{figure}
\centerline{\psfig{file=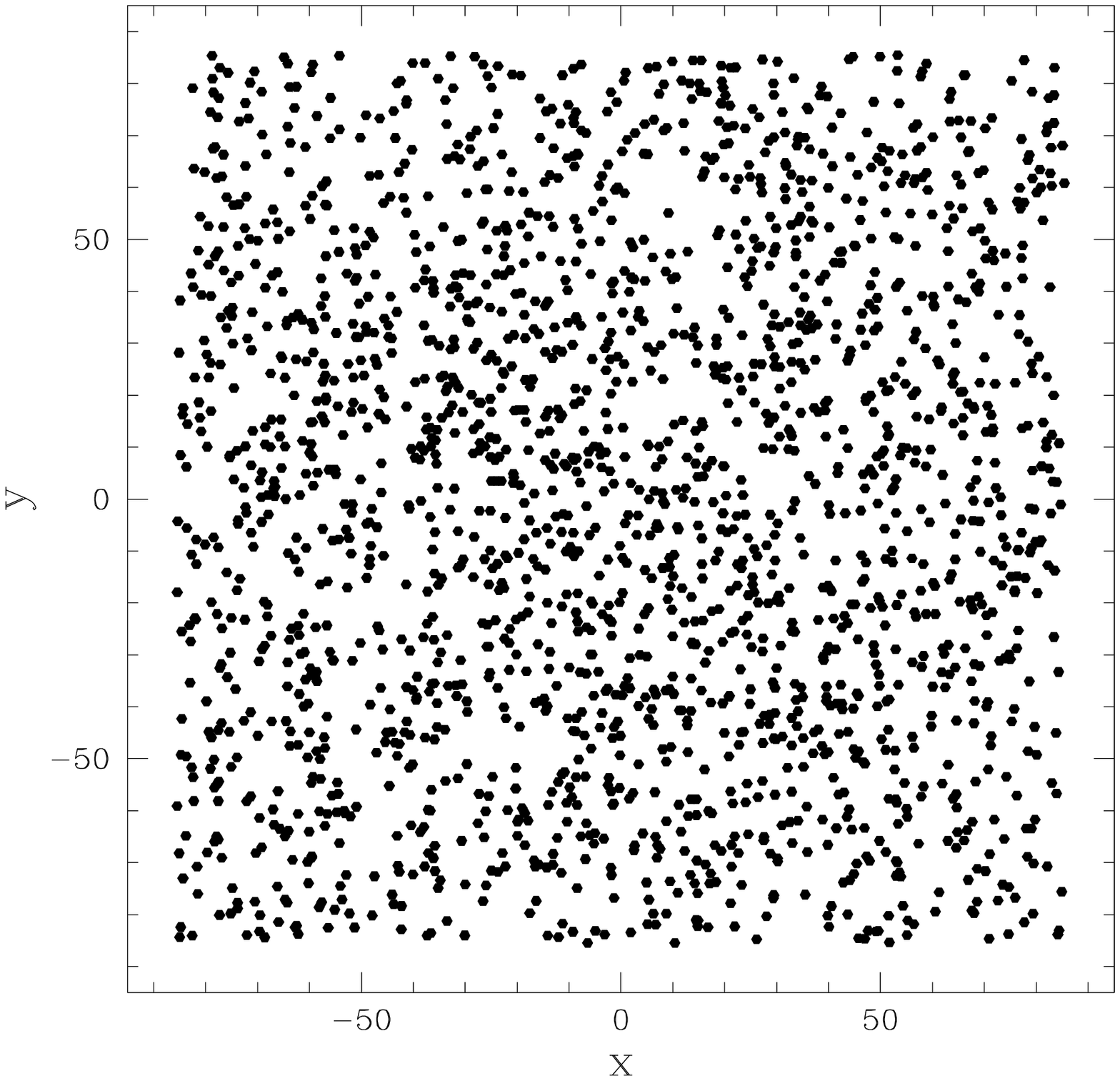,height=8cm,clip=true}}
\caption{Same as Fig.\protect\ref{x-y-core} but with $E/N^{5/3}=0.12$.
Homogeneous phase (see Fig.\protect\ref{fig_TE4}).}
\label{x-y-gas}
\end{figure}

The non-trivial physics behind the cross-over is 
phenomenologically displayed by Figs. \ref{x-y-hom} through \ref{x-y-gas}, 
where a few snapshots of the spatial distribution of the $N$ interacting
masses are projected onto the $x-y$ plane. It turns out that the $V^{-4/3}$
branch of the equation of state corresponds to the clustered phase, whereas 
the same picture obtained in the $V^{-1}$ branch displays an homogeneous
distribution of particles, typical of a gas phase. Figures \ref{x-y-hom} 
and \ref{x-y-clust} refer to the Fourier-truncated model and are obtained
keeping energy constant and varying the volume of the box, whereas Figs.
\ref{x-y-core} through \ref{x-y-gas} refer to the direct $N$-body system and
have been obtained by keeping the volume constant and varying the energy.

A remarkable property of this equation of state  
is the absence of a critical point of flat tangency, i.e. $dP/dV=0$, as
it occurs in the liquid-gas transition described by the van der Waals 
equation. The kink in the van der Waals isotherms below the critical one
(corresponding to the phase coexistence and to the existence of the latent 
heat) is here absent, thus the ``gravitational condensation'' appears rather
different from the gas-liquid condensation. 

A final comment about the energy values of the $P-V$ curves is in order.
At any finite ${\cal N}_w$, the Hamiltonian (\ref{Hbar}) implies a truncation
error in the potential energy with respect to the Hamiltonian (\ref{Hvera}). 
At fixed volume this implies just a shift in 
energy scale, making easy the comparison between the Fourier-truncated model
and the direct model. But when volume is varied, also the truncation error
$\Delta U$ varies. It is numerically confirmed that the expected 
$V$-dependence of the correction to the potential 
energy scale is proportional to $V^{-1/3}$ (for example, at $N=100$, 
${\cal N}_w=7$, $V=10^2$ and $V=2.5\ 10^5$, we found $\Delta U/N^{5/3}=-0.85$ 
and $\Delta U/N^{5/3}=-0.065$ respectively). 
Thus, the truncation can only shift the
value of $V$ at which the crossover occurs, without qualitatively changing
the phenomenology. The larger ${\cal N}_w$ the smaller the shift will be.  

\section{Comparison with theoretical predictions}
\label{theory}
It was suggested by Terletsky (1952) that for a large mass $M$ 
of gas in a volume $V$ and at temperature $T$, composed of $N$ particles
under a boundary pressure $p$, the equation of state of the perfect gas
(Boyle's law) should be corrected to
\begin{equation}
P V = N k_B T - \alpha G M^2 V^{-1/3}~,
\label{bonnor}
\end{equation}
where $\alpha$ accounts for the shape of the mass and $G$ is the gravitational
coupling constant.
Shortly after, it was shown that such a correction of Boyle's law results
in gravitational instability (Bonnor 1956).

Two theoretical milestones followed with the papers of Antonov (1962)  
and Lynden-Bell $\&$ Wood (1968), where a gravitating system of point 
particles in
a spherical box and the thermodynamics  of a self-gravitating isothermal
gas sphere were respectively considered.
Antonov's very interesting result is about the existence of the gravothermal
catastrophe when the density contrast $\varrho_c/\varrho_e$ 
between centre and edge of the box
exceeds the value $709$. Lynden-Bell and Wood (1968) related Antonov's 
result with
the existence of negative heat capacity and found the remarkable result that
stable isothermal spheres with negative specific heat exist in the
density contrast range between $32.2$ and $709$; the specific heat diverges
at $\varrho_c/\varrho_e =32.2$ and becomes positive below this threshold
value; above $\varrho_c/\varrho_e =709$ there is a runaway in the catastrophic
phase. 
Moreover, the minimum attainable temperature by an equilibrium isothermal 
sphere of radius $r_e$ and mass $M$ is 
\begin{equation}
T_{min} = \frac{G m M}{2.52 k_B r_e}
\label{tmin}
\end{equation}
and is achieved for a density contrast $\varrho_c/\varrho_e=32.2$ where
$c_V$ becomes negative.

Our present study naturally represents the microscopic counterpart of the
thermodynamic framework where all the above mentioned results were obtained.

\begin{figure}
\centerline{\psfig{file=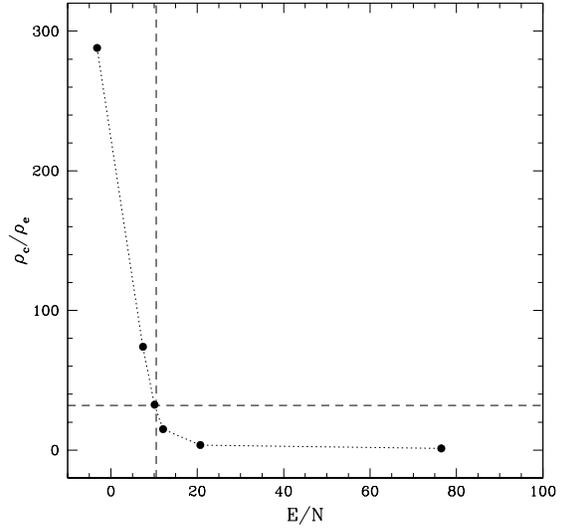,height=8cm,clip=true}}
\caption{Density contrast {\it vs} energy per particle for the 
Fourier-truncated model with $N=100$, ${\cal N}_w=7$ and reflecting 
boundary conditions. The vertical dashed line separates $c_V<0$ (left) from
$c_V>0$ (right). The horizontal dashed line indicates the density contrast 
centre to edge $\varrho_c/\varrho_e=32$. }
\label{dens-contr}
\end{figure}

In the preceding section we have seen that the dynamical-microcanonical
equation of state of a self-gravitating $N$-body system yields a numerical
result in perfect agreement with that of Eq.(\ref{bonnor}), at least as far as
the functional form of the equation of state is concerned, and even though we
obtained isoenergetic $P-V$ curves instead of isothermal ones as is implicit
in Eq.(\ref{bonnor}). The transition point, separating
Boyle's law from the polytropic one, actually corresponds to the loss of 
stability of the homogeneous gas phase which goes into a clustered state, 
as is also 
qualitatively shown by Fig.s \ref{x-y-hom} and \ref{x-y-clust}. Moreover,
we have found that the transition actually occurs at the minimum attainable
temperature\footnote{We have $G=m=k_B=1$, $r_e=L$
and $M=N$. Trivial algebra gives $T_{min}/N^{2/3}=(N/V)^{1/3}/2.52$, whence
$T_{min}/N^{2/3}\simeq 0.4$ for $\varrho =1$, whereas we found $0.6$ -- as
shown in Figure \protect\ref{fig_TE5} -- and $T_{min}/N^{2/3}\simeq 0.03$
for $\varrho =4\ 10^{-4}$, whereas we found $0.04$, as
shown in Figure \protect\ref{fig_TE4}.} in fairly good agreement (taking 
into account the difference of the geometry of the box) with the prediction 
of Eq.(\ref{tmin}), and that below the energy that corresponds to the minimum
attainable temperature the
specific heat is negative, as shown in Section \ref{spec-heat}. 
The increase of the density $N/V$ implies a shift of the transition energy 
density towards more negative values -- as shown by Figs. \ref{fig_TE4} and
\ref{fig_TE5} --  and this is coherent with the Lynden-Bell $\&$ Wood 
prediction of a critical density contrast above which $c_V<0$ and clustering
occurs. In fact, increasing the density amounts to decrease the density 
contrast (if $E/N$ is kept constant) and increasing $E/N$ amounts to decrease 
the density contrast (at constant $\varrho$), thus, in order to keep constant
a given ratio $\varrho_c/\varrho_e$ (for example the threshold value) it is
necessary to decrease $E/N$.

Finally, in
Fig.\ref{dens-contr}, we show that the density contrast between centre and
edge when $c_V$ becomes negative is actually in strikingly good agreement with
the prediction of Lynden-Bell $\&$ Wood. Even if these authors warned that
``both these instabilities\footnote{Occurring at $\varrho_c/\varrho_e=32.2$
and at the Antonov threshold $\varrho_c/\varrho_e=709.$} 
depend on the presence 
of the heat-bath; thermally isolated systems do not suffer this type of 
instability'', we found that the clustering instability shows up also in the
absence of a heat-bath, whereas this is not the case of Antonov's gravothermal
instability.

There are several possible physical explanations for the absence of a 
dynamical
counterpart of the gravothermal catastrophe. A first possibility is that its 
growth rate is so small that a huge time is
required to observe it; this is what has been found for a hybrid model 
-- mean-field with local perturbations -- where the time scale to develop the 
gravothermal catastrophe is larger than the Hubble time (Sygnet et al. 
1984). Another dynamical mechanism that can inhibit the collapse is the 
formation of binary systems (Heggie $\&$ Aarseth 1992).

\section{Conclusions}
\label{concl}
The present paper has mainly dealt with the dynamic (``microscopic'')
origin of  the thermodynamic behaviour of self-gravitating systems and --
in close connection with this topic -- with the problem of their appropriate
statistical mechanical treatment.

The dynamics of self-gravitating $N$-body systems, described by two 
different kinds of regularization of the newtonian interaction, has been
numerically investigated. Dynamics provides the time averages 
of certain quantities which -- with the aid of suitable formulae developed 
in the field of Molecular Dynamics -- yield thermodynamics.
The link between dynamics and thermodynamics is made in the microcanonical
ensemble of statistical mechanics.

The regularizations of the newtonian interaction
are the standard softening and a truncation of the Fourier expansion series 
of the two-body potential. $N$ particles of
equal masses are constrained in a finite three dimensional volume.

The introduction of a Fourier-truncated model makes possible a 
mean-field-like decoupling of 
the degrees of freedom, which is of great prospective theoretical interest 
for a statistical mechanical treatment of self-gravitating  systems. 
Moreover, just through such a decoupling, an order parameter
can be naturally introduced to signal the occurrence of a phase transition. 
Actually, through the computation of the caloric curve, of the specific heat
and of the  order parameter, clear evidence is found for a 
dynamical signature of a phase transition of clustering type which appears
as a {\it second order} phase transition. Thus, the gravitational condensation
seems to take place trough a transition milder than that of a gas-liquid 
condensation, which is a first order transition with a finite jump in the 
entropy. This is also coherent with our numerical computation of
the microcanonical equation of state: the counterpart of the clustering
transition on the $P - V$ plane is a cross-over between a polytrope of index 
$n=3$, i.e. $P=K V^{-4/3}$ and a perfect gas law $P=K^\prime V^{-1}$, without
any kink in the $P~vs~V$ curves as it would be expected for a first order
transition.

The very good agreement between direct simulations of the $N$-body systems 
and Fourier-truncated models is also interesting because it reveals that 
the relevant informations for thermodynamics are mainly conveyed by the large 
spatial scales rather than by the small ones.
This is in agreement with the fact that the statistical mechanical 
peculiarities of $N$-body self-gravitating systems are due to the 
long-range nature of the newtonian interaction and are not influenced 
by its short-scale singularity.

The scale invariance  (Heggie $\&$ Mathieu 1986) of the gravitational 
$N$-body systems is broken by constraining the particles in a box. The
physical reason is that the box introduces a new length scale besides
the gravitational one. The clustering phase transition is a consequence
of the existence of these two independent length scales.
Therefore, the physical meaning of what we found
is that whenever an extra length scale is present besides the gravitational
one, an $N$-body system can undergo a clustering transition. 
The use of the box is perhaps the simplest way of breaking scale invariance,
but a halo of diffuse matter or any other external
potential could in principle work.

We have found that newtonian dynamics naturally yields a number of classical 
results obtained within a phenomenological (thermodynamic) framework.
A remarkable difference exists between the theoretical prediction of two kinds
of transitions, the clustering and the gravothermal catastrophe, and our 
numerical
results: only clustering is recovered in the dynamical approach. Actually, the
possibility of a dynamical suppression of the gravothermal catastrophe 
(motivated either by a huge instability time or by the formation of binaries) 
has been discussed in the more recent literature.

Finally, we have compared the microcanonical (dynamical) averages to their 
canonical ensemble counterparts, obtained through standard Monte Carlo
computations.
A remarkable result found here concerns the ensemble inequivalence. 
This is a-priori expected because of the appearance of negative specific 
heat -- forbidden in the canonical ensemble -- in the dynamically
worked out thermodynamics.
However, the inequivalence is so strong that no signal of the clustering 
transition seems to survive in the canonical ensemble.
In the case of other models some trace is left (Lynden-Bell $\&$ Lynden-Bell 
1977), but perhaps the transition here is too soft.
A reliable statistical mechanical approach to (regularized) self-gravitating 
systems seems possible only in the microcanonical ensemble. 

\section*{Acknowledgments}
It is a pleasure to warmly thank D. Lynden-Bell for a long and very fruitful
discussion with one of us (MP) held at IoA, Cambridge.
We have also profited of helpful discussions with E.G.D. Cohen, D.H.E. Gross, 
S.Ruffo and D. Mukamel.

\appendix
\section[]{}
For the sake of clarity and self-containedness, in this Section we briefly
recall some basic facts about dynamics, statistical
mechanics and thermodynamics. 
\subsection{From dynamics to statistics}

The process of measuring an observable $f(q,p)$, function
of the microscopic coordinates $q$ and momenta $p$ of the particles composing
a macroscopic system, always 
conceptually involves a time average, necessarily carried out 
in a finite time interval $T$, of the form
\beq
\overline{f_T (t)} = \frac{1}{T} \int_t^{t+T} 
f(q(\tau),p(\tau))\,d\tau~.
\label{misura_t}
\eeq
Let us now suppose that the system is at {\em equilibrium}, 
i.e., that the time average (\ref{misura_t}), for 
sufficiently large $T$, does not depend upon $t$, whence we 
can write
\beq
\overline{f} = \lim_{T\to\infty}\overline{f_T (t)}~.
\label{time_average}
\eeq
The founding hypothesis of equilibrium statistical mechanics 
is that (\ref{time_average}) exists and is independent of 
the initial condition $(q(t_0),p(t_0))$, except for a zero 
measure set in the accessible region of phase space.
In the case of an isolated system, such a region is the 
constant-energy hypersurface $\Sigma_E$, and the measure we 
are referring to is the dynamically invariant measure in the 
phase space --- the Liouville measure --- restricted to 
$\Sigma_E$. If this hypothesis, known as the {\em 
ergodic hypothesis}, is valid, then for any observable 
$f(q,p)$ we have
\beq
\overline{f} = \langle f \rangle_\mu~,
\label{ergodic_hypothesis}
\eeq
where $\langle \cdot \rangle_\mu$ denotes the microcanonical 
phase average
\beq
\langle f \rangle_\mu = 
\frac{1}{\omega_N}\int\  dq\ dp\ f(q,p) \ 
\delta[H(q,p) - E]\, ,
\label{media_mu}
\eeq
where $p=\{p_1,\dots , p_N\}$ and $q=\{q_1,\dots , q_N\}$ and $\omega$ is the 
normalization 
\beq
\omega_N(E) =\int \  dq\ dp\ \delta[H(q,p) - E]~.
\label{omega}  
\eeq
 
The ergodic hypothesis makes the problem of the computation 
of (\ref{time_average}) no longer dependent on the knowledge 
of the dynamics. If we consider a non-isolated system,  
in contact with a much larger system, and we apply Eq. 
(\ref{media_mu}) to the sum of the two systems, we obtain the 
canonical formulation of statistical mechanics. As is well known, the 
Gibbs$^\prime$ probability density in phase space is now proportional 
to $\exp(-\beta H)$ rather than to $\delta(H - E)$, 
where $\beta=1/k_BT$ ($k_B$ is the Boltzmann constant and $T$ is the
temperature), so that
\beq
\langle f \rangle_G = 
\frac{1}{Z_N}\int\  dq\ dp\ f(q,p) \ 
\exp[-\beta H(q,p)]\, ,
\label{media_G}
\eeq
where
\beq
Z_N(T,V) =\int \  dq\ dp\ \exp[-\beta H(q,p)]\,
\label{zetaG}  
\eeq
is the {\it canonical partition function}.
\footnote{Szilard (1925) showed that the functional form
of the canonical distribution is constrained by a consistency requirement 
with the Second Law of Thermodynamics.}.

It is natural to think of dynamics as the basic source of 
statistical mechanics, and therefore one can wonder under what conditions
the dynamics is ergodic, so that Eq. (\ref{ergodic_hypothesis}) holds. 
Even if there is no rigorous proof of ergodicity 
but for abstract systems, there are only 
trivial examples of physical systems where it is 
possible to {\em observe} a violation of 
ergodicity (with the exception of amorphous materials, like glasses
or spin-glasses), i.e. where the predictions of 
statistical mechanics are in disagreement with
the outcomes of experimental measurements at equilibrium.

The real problem is the dynamical realization of equilibrium. In statistical 
mechanics it is {\em ab initio} assumed that a system is at equilibrium, 
but the equilibrium concept is very subtle, and depends in a 
crucial way on the {\em time scales} over which a phenomenon 
is observed. In Feynman's words, at equilibrium all the 
``fast'' things have already happened, while the ``slow'' 
ones not yet.
 
This is the reason why from a physical point of view it is 
{\em mixing}, rather than ergodicity, the most important 
feature of a dynamical system. A mixing dynamics is also ergodic;
the converse is not true. Under a mixing dynamics, a generic 
probability distribution in phase space $\varrho(q,p,t)$ 
converges exponentially fast to the stationary equilibrium distribution. 
Therefore mixing is responsible for the relaxation to 
equilibrium, and for the convergence of time averages to statistical 
equilibrium values on finite time scales.  This sheds a new 
light on the role of dynamics in explaining thermodynamical 
phenomena, because there is a tight relationship between mixing and 
that kind of dynamical instability which is called {\it deterministic chaos}.

The reason to believe that chaos
is at the origin of phase space mixing is that in all the systems where the 
mixing property can be rigorously ascertained, mixing  
is found to be a consequence of chaos.

\subsection{From statistics to thermodynamics}

Statistical mechanics also provides a link between the microscopic 
(dynamic) 
and macroscopic (thermodynamic) descriptions of large collections of objects. 
The main frameworks are (Huang 1963) the microcanonical ensemble 
(when energy and number of particles are given), the canonical ensemble 
(when temperature and number of particles are given) and 
grand-canonical ensemble (when total energy and number of particles are 
allowed to fluctuate). 
The equivalence of these ensembles, in the large $N$
limit (thermodynamic limit), is a fundamental point, so that the choice of the
statistical ensemble is only a matter of practical convenience.

In the microcanonical ensemble the link with thermodynamics is made through 
the following definition of entropy ($V$ is the physical volume, $N$ is the
number of particles) 
\beq
S(E, V, N)
 =\ k_B\ \log \ \omega_N(E) .
\label{entropy_surf}
\eeq
Another definition of the microcanonical ensemble volume in phase space,
alternative to $\omega_N$ in Eq. (\ref{omega}), is
\beq
\Omega_N(E) = \int \  dq\ dp \Theta[E - H(q,p)]~,
\label{ball}
\eeq
where $\Theta(\cdot)$ is the Heaviside step function, and the entropy is 
now given by
\beq
\tilde S(E, V, N)
 =\ k_B\ \log \ \Omega_N(E) .
\label{entropy_ball}
\eeq
The two definitions of entropy differ by ${\cal O}(1/N)$ terms and therefore 
coincide in the thermodynamic limit.
With both definitions of entropy, temperature is obtained as 
\beq
\frac{1}{T(E)}=\frac{\partial S}{\partial E}\ .
\label{temp}
\eeq

Strictly speaking, the above definitions of entropy are arbitrary and justified
a-posteriori mainly by showing that they are consistent with the laws of 
thermodynamics (Huang 1963).

In the canonical ensemble the link with thermodynamics is based on the 
following definition of the Helmoltz free energy
\beq
F(T,V,N) = 
- k_B T\ \log \ Z_N(T, V)~,
\label{free_ener}
\eeq
where $Z_N$ is the partition function defined in Eq. (\ref{zetaG}); from the 
function $F$ all the 
other thermodynamic functions are obtained through standard Maxwell's
relations. Also the definition of $F(T,V,N)$ is a-priori arbitrary and is
given validity a-posteriori.

At the macroscopic level, a many-body system has a good thermodynamic behaviour
only if the microscopic interaction potential fulfils two basic properties:
{\it stability} and {\it temperedness} (Ruelle 1969). 
Loosely speaking, temperedness means
that the positive part of the interaction energy between particles at large 
distances is vanishingly small. The negative part of the interaction is left
arbitrary by the definition of temperedness, thus it has to be accompanied by
the stability condition -- which is the relevant condition for gravitational 
interactions --
\beq
U({\bf r}_1,\dots,{\bf r}_N)\geq -N B~,
\label{stability}
\eeq
where $B\geq 0$.
From these conditions the existence of the thermodynamic limit and the
equivalence of the statistical ensembles follow. The existence of the
thermodynamic limit means that by letting $N\rightarrow\infty$ and
$V\rightarrow\infty$ so that $N/V$ remains finite, then also the energy 
density, entropy density and Helmoltz free energy density remain finite.
In other words, energy, entropy, free energy are {\it extensive} quantities.

The {\it ensemble equivalence} holds in the thermodynamic limit and means that
the same thermodynamics can be described by means of microcanonical and 
canonical ensembles, for example.
\section[]{}
Let us briefly sketch here how the computation of the coefficients of the
mean-field Hamiltonian (\ref{Hmf}) should proceed and, similarly, how this
mean-field version of the gravitational $N$-body Hamiltonian (\ref{Hbar})
would make feasible some analytic computation of microcanonical statistical
averages.
The microcanonical average $\langle S_{l,m,n}\rangle$ of any coefficient
\[ 
 S_{l,m,n} =
\sum_{j=1}^N\sin ( k_l x_j) \sin (k_m y_j) \sin (k_n z_j)
\]
is in principle computable according to Eq.(\ref{media_mu}), where $H(p,q)$ 
is given by Eq.(\ref{Hbar}). However, in practice such a computation is 
unfeasible. If the Hamiltonian (\ref{Hbar}) is replaced with the mean-field 
Hamiltonian (\ref{Hmf}), then
\[
\langle S_{l,m,n}\rangle =
\]
\[ 
\frac{1}{\omega_N}\int\prod_{i=1}^N d{\vec x}_i\ d{\vec p}_i\ 
\delta[H_{mf}({\vec p}_k,{\vec x}_k,\langle S_{l,m,n}\rangle) - E]\ 
S_{l,m,n}~.
\]
This integral equation in the unknown $\langle S_{l,m,n}\rangle$ is the
{\it consistency} equation for the mean-field approximation. 

Now the following identity is very useful
\[
\langle S_{l,m,n}\rangle =
\]
\[
{\cal L}^{-1}\left\{ {\cal L}\left[ 
\frac{1}{\omega_N}\int\prod_{i=1}^N d{\vec x}_i\ d{\vec p}_i\ 
\delta[H_{mf}({\vec p}_k,{\vec x}_k,\langle S_{l,m,n}\rangle) - E]
S_{l,m,n}\right]\right\}~,
\]
where ${\cal L}$ stands for Laplace transform and ${\cal L}^{-1}$
for its inverse. In fact, the momenta are easily integrated (Pearson et
al. 1985) leaving
\[
\langle S_{l,m,n}\rangle =
\]
\[
{\cal L}^{-1}\left\{ \frac{1}{\omega_N C_0}\left(\frac{1}{s}\right)^
{\frac{3N}{2}-4}
\right.
\]
\[
\left. 
\prod_{i=1}^N \left[ \int d{\vec x}_i\  
e^{-s U_i({\vec x}_i,\langle S_{l,m,n}\rangle)}\sin ( k_l x_i) \sin (k_m y_i) 
\sin (k_n z_i)
\right]\right\}~,
\]
where $C_0$ is a numerical constant, $s$ is the variable of the Laplace 
transform, and thanks to the decoupling of the degrees of freedom in
$H_{mf}$, all the triple integrals in square brackets are equal and
independent one from another because the potential energy $U$ in Eq.(\ref{Hmf})
is the sum of independent contributions $U_i$. If we denote the generic
triple integral with $F(s,\langle S_{l,m,n}\rangle)$, we are finally left with
\[
\langle S_{l,m,n}\rangle ={\cal L}^{-1}\left\{ \frac{1}{\omega_N C_0}
\left(\frac{1}{s}\right)^{\frac{3N}{2}-4}
\left[ F(s,\langle S_{l,m,n}\rangle)\right]^N \right\}~,
\]
which, in general, can be worked out through some approximate method, the most
popular being the saddle point method (Huang 1963).

Once the coefficients $\langle S_{l,m,n}\rangle$ have been computed, the
mean-field Hamiltonian $H_{mf}$ is completely specified and can be used to
compute other microcanonical averages through the same guideline depicted 
above.

\newpage

\bsp

\label{lastpage}

\end{document}